\documentclass[10pt,conference,letterpaper]{IEEEtran}

\usepackage{graphicx}
\usepackage{amsmath,amssymb,epsfig,latexsym,amssymb,graphics}
\usepackage[numbers,sort&compress]{natbib}

\usepackage[normalem]{ulem}
\usepackage{url}
\usepackage{epsfig}
\usepackage{xspace}
\usepackage{epstopdf}
\usepackage{color}
\usepackage{pstricks}
\usepackage{times}
\usepackage{epsfig}
\usepackage{graphicx}
\usepackage{xspace}
\usepackage{subfigure}
\usepackage{float}
\usepackage{latexsym}
\usepackage{algorithm}
\usepackage[noend]{algorithmic}
\usepackage{url}
\usepackage{color}
\usepackage{colortbl}
\usepackage{hyperref}
\usepackage{fancynum}
\usepackage{xcolor,colortbl}
\usepackage{caption}
\usepackage{textcomp}
\usepackage{mathrsfs}
\usepackage{tikz}
\usepackage{amsmath}

\newcommand{\topcaption}{%
 \setlength{\belowcaptionskip}{1pt}%
 \caption}

\definecolor{Gray}{gray}{0.85}
\definecolor{LightCyan}{rgb}{0.88,1,1}
\definecolor{Yellow}{rgb}{1,1,0.5}
\definecolor{Peachpuff}{rgb}{1,0.8,0.6}
\definecolor{Lavender}{rgb}{0.91,0.91,0.98}

\newcolumntype{a}{>{\columncolor{Gray}}c}
\newcolumntype{b}{>{\columncolor{white}}c}

\newcommand{\cg}{\cellcolor{Gray}}

\newcommand{\cy}{\cellcolor{Yellow}}
\newcommand{\cp}{\cellcolor{Peachpuff}}

\newcounter{example}[section]
\renewcommand{\theexample}{\nthesection.\arabic{example}}
\newenvironment{example}{
     \refstepcounter{example}
     {\vspace{0ex} \noindent\bf  Example  \theexample:}}{
     \eop\vspace{0ex}} 

\newcounter{definition}[section]
\renewcommand{\thedefinition}{\nthesection.\arabic{definition}}
\newenvironment{definition}{
     \refstepcounter{definition}
     {\vspace{0ex} \noindent\bf  Definition  \thedefinition:}}{
     \eop\vspace{0ex}} 

\newcounter{theorem}[section]
\renewcommand{\thetheorem}{\nthesection.\arabic{theorem}}
\newenvironment{theorem}{\begin{em}
        \refstepcounter{theorem}
        {\vspace{0ex} \noindent\bf  Theorem  \thetheorem:}}{
        \end{em}\eop\vspace{0ex}} 

\newcounter{lemma}[section]
\renewcommand{\thelemma}{\nthesection.\arabic{lemma}}
\newenvironment{lemma}{\begin{em}
        \refstepcounter{lemma}
        {\vspace{0ex}\noindent\bf Lemma \thelemma:}}{
        \end{em}\eop\vspace{0ex}} 

\newcounter{remark}[section]
\renewcommand{\theremark}{\nthesection.\arabic{remark}}

\newcounter{property}[section]
\renewcommand{\theproperty}{\nthesection.\arabic{property}}
\newenvironment{property}{\begin{em}
        \refstepcounter{property}
        {\vspace{0ex}\noindent\bf Property \theproperty:}}{
        \end{em}\eop\vspace{0ex}} 
        
\newcounter{observation}[section]
\renewcommand{\theobservation}{\nthesection.\arabic{observation}}

\newcommand{\kw}[1]{{\ensuremath {\mathsf{#1}}}\xspace}
\newcommand{\kwnospace}[1]{{\ensuremath {\mathsf{#1}}}}

\newcommand{\myitem}{\noindent$\bullet$\xspace}

\newcommand{\stitle}[1]{\vspace{0ex} \noindent{\bf #1}}
\newcommand{\sstitle}[1]{\vspace{0ex} \noindent{\it #1}}

\newcommand{\nthesection}{\arabic{section}}
\newcommand{\eop}{\hspace*{\fill}\mbox{$\Box$}}

\newcommand{\reffig}[1]{Fig.~\ref{fig:#1}}
\newcommand{\refsec}[1]{Section~\ref{sec:#1}}
\newcommand{\refsubsec}[1]{Subsection~\ref{subsec:#1}}
\newcommand{\reftab}[1]{Table~\ref{tab:#1}}
\newcommand{\refalg}[1]{Algorithm~\ref{alg:#1}}
\newcommand{\refeq}[1]{Eq.~\ref{eq:#1}}
\newcommand{\refdef}[1]{Definiton~\ref{def:#1}}
\newcommand{\refthm}[1]{Theorem~\ref{thm:#1}}
\newcommand{\reflem}[1]{Lemma~\ref{lem:#1}}
\newcommand{\refex}[1]{Example~\ref{ex:#1}}

\newcommand{\refpr}[1]{Property~\ref{pr:#1}}

\newcommand{\myproof}{\noindent{\textit{Proof: }}}

\newcommand{\nb}{\kw{nbr}}
\newcommand{\dg}{\kw{deg}}
\newcommand{\core}{\kw{core}}
\newcommand{\coreub}{\overline{\kw{core}}}
\newcommand{\cub}{{\kw{ub}}}
\newcommand{\kmax}{k_{max}}
\newcommand{\emcore}{\kw{EMCore}}

\newcommand{\imcore}{\kw{IMCore}}
\newcommand{\semicore}{\kw{SemiCore}}
\newcommand{\semicorep}{\kw{SemiCore}^+\xspace}
\newcommand{\semicores}{\kw{SemiCore}^*\xspace}
\newcommand{\semiadd}{\kw{SemiInsert}}
\newcommand{\semiadds}{\kw{SemiInsert}^*\xspace}

\newcommand{\semidels}{\kw{SemiDelete}^*\xspace}
\newcommand{\imadd}{\kw{IMInsert}}
\newcommand{\imdel}{\kw{IMDelete}}
\newcommand{\update}{\kw{update}}
\newcommand{\localcore}{\kw{LocalCore}}
\newcommand{\computecnt}{\kw{ComputeCnt}}
\newcommand{\computecnts}{\kw{ComputeCnt}^*}
\newcommand{\updatenbrcnt}{\kw{UpdateNbrCnt}}
\newcommand{\cnt}{\kw{cnt}}
\newcommand{\cnts}{\kw{cnt}^*\hspace*{-0.03cm}}
\newcommand{\num}{\kw{num}}
\newcommand{\act}{\kw{active}}
\newcommand{\corg}{\kw{c}_{\kw{old}}}
\newcommand{\status}{\kw{status}}
\newcommand{\updaterange}{\kw{UpdateRange}}

\newcommand*\circled[1]{{\tikz[baseline=(C.base)]\node[fill=gray!40,draw,circle,inner sep=0.1pt,line width=0.2mm](C) {\small #1};}}
\newcommand*\circledq[1]{{\tikz[baseline=(C.base)]\node[fill=gray!40,draw,circle,inner sep=0.8pt,line width=0.2mm](C) {\small #1};}}

\newcommand{\statusn}{\circled{$\times$}\xspace}

\newcommand{\statuse}{\circled{$\phi$}\xspace}

\newcommand{\statusq}{\circledq{$?$}\xspace}

\newcommand{\statusy}{\circled{{\scriptsize$\surd$}}\xspace}

\newcommand{\dblp}{\kwnospace{DBLP}\xspace}
\newcommand{\wiki}{\kwnospace{WIKI}\xspace}
\newcommand{\lj}{\kwnospace{LJ}\xspace}
\newcommand{\orkut}{\kwnospace{Orkut}\xspace}
\newcommand{\ytb}{\kwnospace{Youtube}\xspace}
\newcommand{\cpt}{\kwnospace{CPT}\xspace}

\newcommand{\twitter}{\kwnospace{Twitter}\xspace}
\newcommand{\uk}{\kwnospace{UK}\xspace}
\newcommand{\sk}{\kwnospace{SK}\xspace}
\newcommand{\clueweb}{\kwnospace{Clueweb}\xspace}
\newcommand{\webbase}{\kwnospace{Webbase}\xspace}
\newcommand{\itc}{\kwnospace{IT}\xspace}
\newcommand{\facebook}{\kwnospace{Facebook}\xspace}

\title{I/O Efficient Core Graph Decomposition at \\Web Scale}

\author{
Dong Wen$^{\natural}$, Lu Qin$^{\natural}$,   Ying Zhang$^{\natural}$, Xuemin Lin$^{\ddag}$, and Jeffrey Xu Yu$^{\S}$ \vspace{1mm}\\
\fontsize{9}{9}\selectfont\itshape

$^{\natural}$Centre for Quantum Computation \& Intelligent Systems, University of Technology, Sydney, Australia \\
$^{\ddag}$The University of New South Wales, Australia \\
$^{\S}$The Chinese University of Hong Kong, China \\
\fontsize{9}{9}\selectfont\ttfamily\upshape
$^{\natural}$dong.wen@student.uts.edu.au; \{lu.qin, ying.zhang\}@uts.edu.au; \\
$^{\ddag}$lxue@cse.unsw.edu.au; $^{\S}$yu@se.cuhk.edu.hk 
}

\begin{document}

\maketitle


\begin{abstract}
Core decomposition is a fundamental graph problem with a large number of applications. Most existing approaches for core decomposition assume that the graph is kept in memory of a machine. Nevertheless, many real-world graphs are big and may not reside in memory. In the literature, there is only one work for I/O efficient core decomposition that avoids loading the whole graph in memory. However, this approach is not scalable to handle big graphs because it cannot bound the memory size and may load most parts of the graph in memory. In addition, this approach can hardly handle graph updates. In this paper, we study I/O efficient core decomposition following a semi-external model, which only allows node information to be loaded in memory. This model works well in many web-scale graphs. We propose a semi-external algorithm and two optimized algorithms for I/O efficient core decomposition using very simple structures and data access model. To handle dynamic graph updates, we show that our algorithm can be naturally extended to handle edge deletion. We also propose an I/O efficient core maintenance algorithm to handle edge insertion, and an improved algorithm to further reduce I/O and CPU cost by investigating some new graph properties. We conduct extensive experiments on $12$ real large graphs. Our optimal algorithm significantly outperform the existing I/O efficient algorithm in terms of both processing time and memory consumption. In many memory-resident graphs, our algorithms for both core decomposition and maintenance can even outperform the in-memory algorithm due to the simple structures and data access model used. Our algorithms are very scalable to handle web-scale graphs. As an example, we are the first to handle a web graph with $978.5$ million nodes and $42.6$ billion edges using less than $4.2$ GB memory. 
\end{abstract}


\vspace*{-0.2cm}
\section{Introduction}
\label{sec:introduction}
Graphs have been widely used to represent the relationships of entities in a large spectrum of applications such as social networks, web search,  collaboration networks, and biology. With the proliferation of graph applications, research efforts have been devoted to many fundamental problems in managing and analyzing graph data. Among them, the problem of computing the $k$-core of a graph has been recently studied \cite{Cheng2011,Sariyuce2013,LiYM14, Cui2014}. Here, given a graph $G$, the $k$-core of $G$ is the largest subgraph of $G$ such that all the nodes in the subgraph have a degree of at least $k$ \cite{seidman1983}. For each node $v$ in $G$, the core number of $v$ denotes the largest $k$ such that $v$ is contained in a $k$-core. The core decomposition problem computes the core numbers for all nodes in $G$. Given the core decomposition of a graph $G$, the $k$-core of $G$ for all possible $k$ values can be easily obtained. There is a linear time in-memory algorithm, devised by Batagelj and Zaversnik \cite{Batagelj2003}, to compute core numbers of all nodes.

\stitle{Applications}. Core decomposition is widely adopted in many real-world applications, such as community detection \cite{Giatsidis2011,Cui2014}, network clustering \cite{verma2012}, network topology analysis \cite{seidman1983, alvarez2006}, network visualization \cite{alvarez2005, kalvarez2005}, protein-protein network analysis \cite{Altaf2003, bader2003}, and system structure analysis \cite{zhang2010}. In addition, many researches are devoted on the core decomposition for specific kinds of networks \cite{dorogovtsev2006, Luczak1991, Pittel1996, Molloy2005, Janson2007, BonchiGKV14}.  Moreover, due to the elegant structural property of a $k$-core and the linear solution for core decomposition, a large number of graph problems use core decomposition as a subroutine or a preprocessing step, such as clique finding \cite{balasundaram2011}, dense subgraph discovery \cite{andersen2009, QinLCZ15},  approximation of betweeness scores \cite{healy2008}, and some variants of community search problems \cite{SozioG10, LiQYM15}. 

\stitle{Motivation}. Despite the large amount of applications for core decomposition in various networks, most of the solutions for core decomposition assume that the graph is resident in the main memory of a machine. Nevertheless, many real-world graphs are big and may not reside entirely in the main memory. For example, the \facebook social network contains $1.32$ billion nodes and $140$ billion edges\footnote{\url{http://newsroom.fb.com/company-info}}; and a sub-domain of the web graph \clueweb contains $978.5$ million nodes and $42.6$ billion edges\footnote{\url{http://law.di.unimi.it/datasets.php}}. In the literature, the only solution to study I/O efficient core decomposition is \emcore proposed by  Cheng et al. \cite{Cheng2011}, which allows the graph to be partially loaded in the main memory.  \emcore adopts a graph partition based approach and partitions are loaded into main memory whenever necessary. However, \emcore cannot bound the size of the memory and to process  many real-world graphs, \emcore still loads most edges of the graph in the main memory. This makes \emcore unscalable to handle web-scale graphs. In addition, many real-world graphs are usually dynamically updating. The complex structure used in \emcore makes it very difficult to handle graph updates incrementally. 

\stitle{Our Solution}. In this paper, we address the drawbacks of the existing solutions for core decomposition and propose new algorithms for core decomposition with guaranteed memory bound. Specifically, we adopt a semi-external model. It assumes that the nodes of the graph, each of which is associated with a small constant amount of information, can be loaded in main memory while the edges are stored on disk. We find that this assumption is practical in a large number of real-world web-scale graphs, and widely adopted to handle other graph problems \cite{zhiwei2013,zhiwei2015,liu2015}. Based on such an assumption, we are able to handle core decomposition I/O efficiently using very simple structures and data access mechanism. These  enable our algorithm to efficiently handle graph updates incrementally under the semi-external model. 

\stitle{Contributions}. We make the following contributions:

\sstitle{(1) The first I/O efficient core decomposition algorithm with memory guarantee}. We propose an I/O efficient core decomposition algorithm following the semi-external model. Our algorithm only keeps the core numbers of nodes in memory and updates the core numbers iteratively until convergency. In each iteration, we only require sequential scans of edges on disk. To the best of our knowledge, this is the first work for I/O efficient core decomposition with memory guarantee. 

\sstitle{(2) Several optimization strategies to largely reduce the I/O and CPU cost}. Through further analysis, we observe that when the number of iterations increases, only a very small proportion of nodes have their core numbers updated in each iteration, and thus a total scan of all edges on disk in each iteration will result in a large number of waste I/O and CPU cost. Therefore, we propose optimization strategies to reduce these cost. Our first strategy is based on the observation that the update of core number for a node should be triggered by the update of core number for at least one of its neighbors in the graph. Our second strategy further maintains more node information. As a result, we can completely avoid waste I/Os and core number computations, in the sense that each I/O is used in a core number computation that is guaranteed to update the core number of the corresponding node. Both optimization strategies can be easily adapted in our algorithm framework.  

\sstitle{(3) The first I/O efficient core decomposition algorithm to handle graph updates}. We consider dynamical graphs with edge deletion and insertion. Our semi-external algorithm can naturally support edge deletion with a simple algorithm modification. For edge insertion, we first utilize some graph properties adopted in existing in-memory algorithms \cite{Sariyuce2013,LiYM14} to handle graph updates for core decomposition. We propose a  two-phase semi-external algorithm  to handle edge insertion using these graph properties. We further explore some new graph properties, and propose a new one-phase semi-external algorithm to largely reduce the I/O and CPU cost for edge insertion. To the best of our knowledge, this is the first work for I/O efficient core maintenance on dynamic graphs. 

\sstitle{(4) Extensive performance studies}. We conduct extensive performance studies using  $12$ real graphs with various graph properties to demonstrate the efficiency of our algorithms. We compare our algorithm, for memory-resident graphs, with \emcore \cite{Cheng2011} and the in-memory algorithm \cite{Batagelj2003}. Both our core decomposition and core maintenance algorithms are much faster and use much less memory than \emcore. In many datasets, our algorithms for core decomposition and maintenance are even faster than the in-memory algorithm due to the simple structure and data access model used. Our algorithms are very scalable to handle web-scale graphs. For instance, we consume less than $4.2$ GB memory to handle the web-graph \clueweb with  $978.5$ million nodes and $42.6$ billion edges. 

\stitle{Outline}. \refsec{problem} provides the preliminaries and problem statement. \refsec{existing} introduces some existing solutions for core decomposition under different settings. \refsec{decomposition} presents our semi-external core decomposition algorithm and explores some optimization strategies to reduce I/O and CPU cost. \refsec{maintenance} discusses how to design semi-external algorithms to maintain core numbers incrementally when the graph is dynamically updated, and investigates some new graph properties to improve the algorithm when handling edge insertion. \refsec{experiment} evaluates all the introduced algorithms using extensive experiments. \refsec{relatedwork} reviews the related work and \refsec{conclusion} concludes the paper.


\vspace*{-0.2cm}
\section{Problem Statement}
\label{sec:problem}
Consider an undirected and unweighted graph $G=(V,E)$, where $V(G)$ represents the set of nodes and $E(G)$ represents the set of edges in $G$. We denote the number of nodes and the number of edges of $G$ by $n$ and $m$ respectively. We use $\nb(u,G)$ to denote the set of neighbors of $u$ for each node $u\in V(G)$, i.e., $\nb(u,G)=\{v|(u,v)\in E(G)\}$. The degree of a node $u\in V(G)$, denoted by $\dg(u,G)$, is the number of neighbors of $u$ in $G$, i.e., $\dg(u,G)=|\nb(u,G)|$. For simplicity, we use $\nb(u)$ and $\dg(u)$ to denote $\nb(u,G)$ and $\dg(u,G)$ respectively if the context is self-evident. A graph $G'$ is a subgraph of $G$, denoted by $G'\subseteq G$, if $V(G')\subseteq V(G)$ and $E(G')\subseteq E(G)$.  Given a set of nodes $V_c\subseteq V$, the induced subgraph of $V_c$, denoted by $G(V_c)$, is a subgraph of $G$ such that $G(V_c)=(V_c,\{(u,v)\in E(G)| u, v \in V_c\})$. 

\begin{definition} \textbf{($k$-Core)} Given a graph $G$ and an integer $k$, the $k$-core of graph $G$, denoted by $G_k$, is a maximal subgraph of $G$ in which every node has a degree of at least $k$, i.e., $\forall v\in V(G_k), d(v,G_k)\geq k$  \cite{seidman1983}. 
\end{definition}

Let $\kmax$ be the maximum possible $k$ value such that a $k$-core of $G$ exists. According to \cite{Batagelj2003}, the $k$-cores of graph $G$ for all $1\leq k\leq \kmax$ have the following property:

\begin{property}
\label{pr:kcore}
$\forall 1\leq k < \kmax: G_{k+1}\subseteq G_k$.
\end{property}

Next, we define the core number for each $v\in V(G)$.

\begin{definition}
\label{def:corenumber}
\textbf{(Core Number)} Given a graph $G$, for each node $v\in V(G)$, the core number of $v$, denoted by $\core(v,G)$, is the largest $k$, such that $v$ is contained in a $k$-core, i.e., $\core(v,G)=\max\{k|v\in V(G_k)\}$. For simplicity, we use $\core(v)$ to denote $\core(v,G)$ if the context is self-evident.
\end{definition}

Based on \refpr{kcore} and \refdef{corenumber}, we can easily derive the following lemma:

\begin{lemma}
\label{lem:kcore}
Given a graph $G$ and an integer $k$, let $V_k=\{v\in V(G)|\core(v)\geq k\}$, we have $G_k = G(V_k)$. 
\end{lemma}

\stitle{Problem Statement}. In this paper, we study the problem of Core Graph Decomposition (or Core Decomposition for short), which is defined as follows: Given a graph $G$, core decomposition computes the the $k$-cores of $G$ for all $1\leq k \leq \kmax$. We also consider how to update the $k$-cores for of $G$ for all $1\leq k\leq \kmax$ incrementally when $G$ is dynamically updated by insertion and deletion of edges. 

According to \reflem{kcore}, core decomposition is equivalent to computing $\core(v)$ for all $v\in V(G)$. Therefore, in this paper, we study how to compute $\core(v)$ for all $v\in V(G)$ and how to maintain them incrementally when graph is dynamically updating . 

Considering that many real-world graphs are huge and cannot entirely reside in main memory, we aim to design I/O efficient algorithms to compute and maintain the core numbers of all nodes in the graph $G$.   To analyze the algorithm, we use the external memory model introduced in \cite{Aggarwal1988}. Let $M$ be the size of main memory and let  $B$ be the disk block size. A read I/O will load one block of size $B$ from disk into main memory, and a write I/O will write one block of size $B$ from the main memory  into disk.

\stitle{Assumption}. In this paper, we follow a semi-external model by assuming that the nodes can be loaded in main memory while the edges are stored on disk, i.e., we assume that $M\geq c\times |V(G)|$ where $c$ is a small constant. This assumption is practical because in most social networks and web graphs, the number of edges is much larger than the number of nodes. For example, in SNAP\footnote{\url{http://snap.stanford.edu/data/}}, among $79$ real-world graphs, the largest graph contains $65$ M nodes and $1.8$ G edges. In KONET\footnote{\url{http://konect.uni-koblenz.de/networks/}},  among $239$ real-world graphs, the largest graph contains $68$ M nodes and $2.6$ G edges. In WebGraph\footnote{\url{http://law.di.unimi.it/}}, among $75$ real-world graphs, the largest graph contains $721$ M nodes and $137.3$ G edges, and the second largest graph contains $978$ M nodes and $42.6$ G edges. In our proposed algorithm of this paper, when handling the two largest graphs in WebGraph, we only require $3.1$ GB and $4.2$ GB memory respectively, which is affordable even by a normal PC.

\stitle{Graph Storage}. In this paper, we use an edge table on disk to store the edges of $G$. e.g., we store $\nb(v_1)$, $\nb(v_2)$, $\ldots$, $\nb(v_n)$ consecutively as adjacency lists in the edge table. We also use a node table on disk to store the offsets and degrees for $v_1$, $v_2$, $\ldots$, $v_n$ consecutively. To load the neighbors of a certain node $v_i\in V(G)$, we can access the node table to get the offset and $\dg(v_i)$ for $v_i$, and then access the edge table to load $\nb(v_i)$.

\vspace*{-0.2cm}
\begin{figure}[h]
\begin{center}
\includegraphics[width=0.45\hsize]{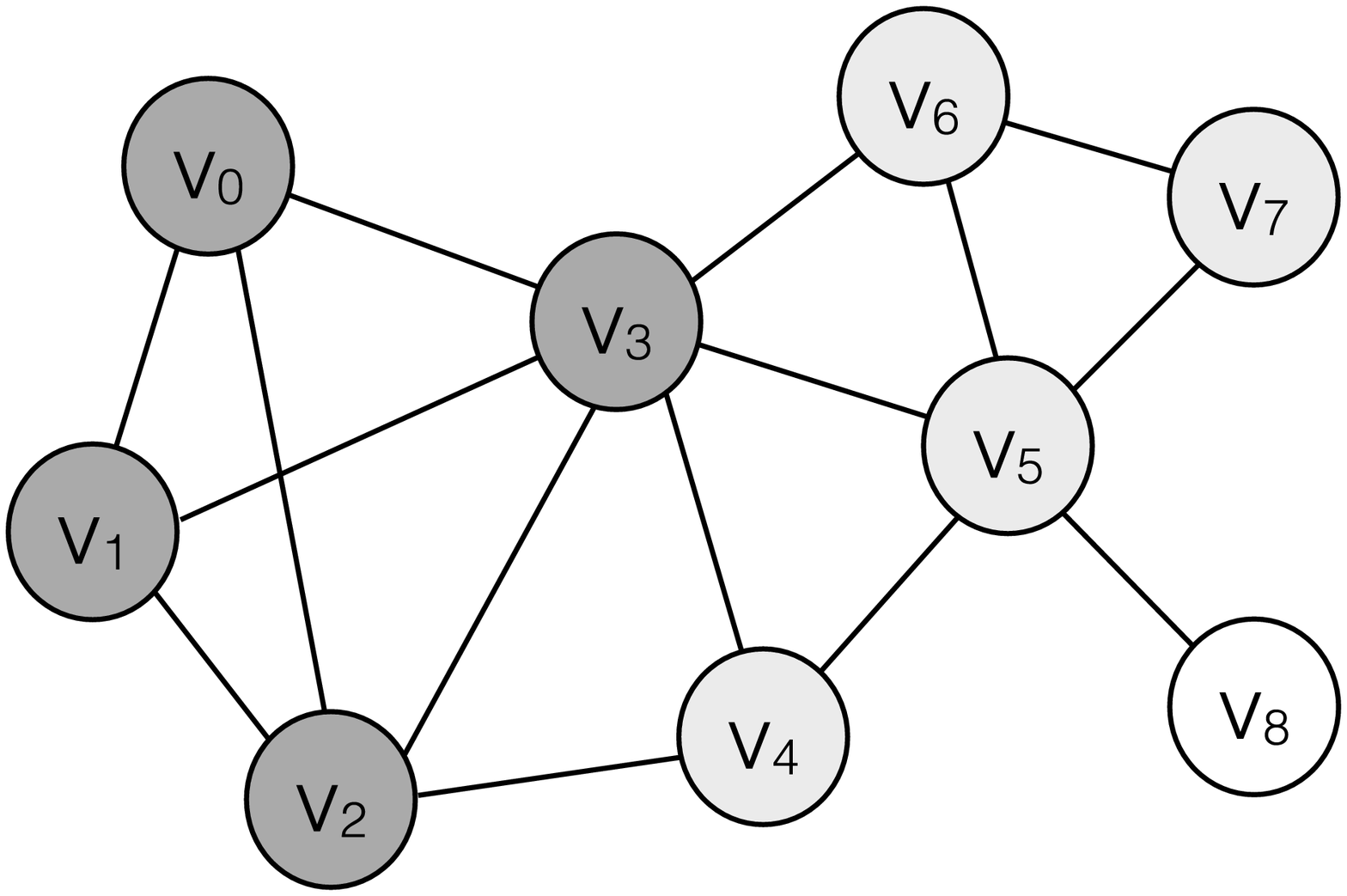}
\vspace*{-0.1cm}
\topcaption{A Sample Graph $G$ and its Core Decomposition}
\vspace*{-0.3cm}
\label{fig:core}
\end{center}
\end{figure}

\begin{example}\label{coreDefinition}
Consider a graph $G$ in \reffig{core}, the induced subgraph of $\{v_0,$ $v_1,$ $v_2,$ $v_3\}$ is a $3\text{-}core$ in which every node has a degree at least $3$. Since no $4$-core exists in $G$, we have $\core(v_0)=\core(v_1)=\core(v_2)=\core(v_3)=3$. Similarly, we can derive that $\core(v_4)=\core(v_5)=\core(v_6)=\core(v_7)=2$ and $\core(v_8)=1$. When an edge $(v_7,v_8)$ is inserted in $G$, $\core(v_8)$ increases from $1$ to $2$, and the core numbers of other nodes keep unchanged. 
\end{example}

\vspace*{-0.2cm}
\section{Existing Solutions}
\label{sec:existing}
In this section, we introduce three state-of-the-art existing solutions for core decomposition in different settings, namely, in-memory core decomposition, I/O efficient core decomposition, and  in-memory core maintenance. 

\stitle{In-memory Core Decomposition}. The state-of-the-art in-memory core decomposition algorithm, denote by $\imcore$,  is proposed in \cite{Batagelj2003}. The pseudocode of $\imcore$ is shown in \refalg{imcore}. The algorithm processes the node with core number $k$ in increasing order of $k$. Each time, $k$ is selected as the minimum degree of current nodes in the graph (line~3). Whenever there exists a node $v$ with degree no larger than $k$ in the graph (line~4), we can guarantee that the core number of $v$ is $k$ (line~5) and we remove $v$ with all its incident edges from the graph (line~6). Finally, the core number of all nodes are returned (line~7). With the help of bin sort to maintain the minimum degree of the graph, $\imcore$ can achieve a time complexity of $O(m+n)$,  which is optimal. 

\begin{algorithm}[t]
\caption{\small{$\imcore($Graph $G)$}}
\label{alg:imcore}
\footnotesize
\begin{algorithmic}[1]
\STATE $G'\leftarrow G$;
\WHILE{$G' \neq \emptyset$}
\STATE $k\leftarrow \min_{v\in V(G')}\dg(v, G')$;
	\WHILE{$\exists v\in V(G'): \dg(v, G')\leq k$}
		\STATE $\core(v)\leftarrow k$;
		\STATE remove $v$ and its incident edges from $G'$;
	\ENDWHILE
\ENDWHILE
\STATE \textbf{return} $\core(v)$ for all $v\in V(G)$;
\end{algorithmic}
\end{algorithm}

\stitle{I/O Efficient Core Decomposition}. The state-of-the-art efficient core decomposition algorithm is proposed in \cite{Cheng2011}. The algorithm, denoted as $\emcore$, is shown in \refalg{emcore}. It first divides the whole graph $G$ into partitions on disk (line~1). Each partition contains a disjoint set of nodes along with their incident edges. An upper bound of $\core(v)$, denoted by $\cub(v)$, is computed for each node $v$ in each partition $P_i$. Then the algorithm iteratively computes the core numbers for nodes in a top-down manner. 

In iteration, the nodes with core values falling in a certain range $[k_l, k_u]$ is computed (line~6-14). Here, $k_l$ is estimated based on the number of partitions that can be loaded in main memory (line~6). In line~7, the algorithm computes the set of partitions each of which contains at leat one node $v$ with $\cub(v)$ falling in $[k_l,k_u]$, and in line~8, all such partitions are loaded in main memory to form an in-memory graph $G_{mem}$. In line~9, an in-memory core decomposition algorithm is applied on $G_{mem}$, and those nodes in $G_{mem}$ with core numbers falling in $[k_l,k_u]$ get their exact core numbers in $G$. After that, for all partitions loaded in memory (line~10), those nodes with exact core numbers computed are removed from the partition (line~11), and the their core number upper bounds and degrees are updated accordingly (line~12). Here the new node degrees have to consider the deposited degrees from the removed nodes. Finally, the in-memory partitions are merged and written back to disk (line~13), and $k_u$ is set to be $k_l-1$ to process the next range of $k$ values in the next iteration.

The I/O complexity of $\emcore$ is $O(\frac{\kmax\cdot (m+n)}{B})$. The CPU complexity of $\emcore$ is $O(\kmax\cdot(m+n))$. However, the space complexity of $\emcore$ cannot be well bounded. In the worst case, it still requires $O(m+n)$ memory space to load the whole graph into main memory. Therefore, $\emcore$ is not scalable to handle large-sized graphs. 

\begin{algorithm}[t]
\caption{\small{$\emcore($Graph $G$ on Disk$)$}}
\label{alg:emcore}
\footnotesize
\begin{algorithmic}[1]
\STATE divide $G$ into partitions ${\cal P}=\{P_1, P_2, \ldots, P_t\}$ on disk;
\FORALL{partition $P_i\in {\cal P}$}
	\STATE compute $\cub(v)$ for all $v\in V(P_i)$;
\ENDFOR
\STATE $k_u\leftarrow +\infty$;
\WHILE{$k_u>0$}
	\STATE estimate $k_l$;
	\STATE ${\cal P}_{mem}\leftarrow \{P_i\in {\cal P}|\exists v\in V(P_i):\cub(v)\in [k_l, k_u]\}$;
	\STATE $G_{mem}\leftarrow$ load partitions in ${\cal P}_{mem}$ in main memory;
	\STATE $\core(v)\leftarrow \core(v,G_{mem})$ for all $\core(v,G_{mem})\in [k_l, k_u]$;
	\FORALL{partition $P_i\in {\cal P}_{mem}$ }
		\STATE remove nodes $v$ with $\core(v,G_{mem})\in [k_l, k_u]$ from $P_i$;
		\STATE update $\cub(v)$ and $\dg(v)$ for all $v\in V(P_i)$;
		\STATE write $P_i$ back to disk (merge small partitions if necessary); 
	\ENDFOR
	\STATE $k_u\leftarrow k_l-1$;
\ENDWHILE
\STATE \textbf{return} $\core(v)$ for all $v\in V(G)$;
\end{algorithmic}
\end{algorithm}

\stitle{In-memory Core Maintenance}. To handle the case when the graph is dynamically updated by insertion and deletion of edges, the state-of-the-art core maintenance algorithms are proposed in  \cite{Sariyuce2013} and \cite{LiYM14}, which are based on the same findings shown in the following theorems:

\begin{theorem}
\label{thm:change1}
If an edge is inserted into (deleted from) graph $G$, the core number $\core(v)$ for any $v\in V(G)$ may increase (decrease) by at most $1$.
\end{theorem}

\begin{theorem}
\label{thm:change2}
If an edge $(u,v)$ is inserted into (deleted from) graph $G$,  suppose $\core(v)\leq \core(u)$ and let $V'$ be the set of nodes whose core numbers have changed,  if $V'\neq \emptyset$, we have:\\
\myitem $G(V')$ is a connected subgraph of $G$;\\
\myitem $v\in V'$; and\\
\myitem $\forall v'\in V': \core(v')=\core(v)$; 
\end{theorem}

Based on \refthm{change1} and \refthm{change2}, after an edge $(u,v)$ is inserted into (deleted from) graph $G$, suppose $\core(v)\leq \core(u)$, instead of computing the core numbers for all nodes in $G$ from scratch, we can restraint the core computation within a small range of nodes  $V'$ in $G$. Specifically, we can follow a two-step approach: In the first step, we can perform a depth-first-search from node $v$ in $G$ to compute all nodes $v'$ with $\core(v')=\core(v)$ and are reachable from $v$ via a path that consists of nodes with core numbers equal to $\core(v)$. Such nodes form a set $V'$ which is usually much smaller than $V(G)$.  In the second step, we only restraint the core number updates within the subgraph $G(V')$ in memory, and each update increases (decreases) the core number of a node by at most $1$. The algorithm details and other optimization techniques can be found in  \cite{Sariyuce2013} and \cite{LiYM14}. 

\vspace*{-0.2cm}
\section{I/O Efficient Core Decomposition}
\label{sec:decomposition}
In this section, we present our basic semi-external algorithm and then discuss how to improve the algorithm by partial node computation. Finally, we will propose an algorithm by eliminating all useless node computations. 

\vspace*{-0.2cm}
\subsection{Basic Semi-external Algorithm}
\label{sec:decomposition:basic}

\stitle{Drawback of \emcore}. \emcore (\refalg{emcore}) is the state-of-the-art I/O efficient core decomposition algorithm. However, \emcore cannot be used to handle big graphs, since the number of partitions to be loaded into main memory in each iteration cannot be well-bounded. In line~7-8 of \refalg{emcore}, as long as a partition contains a node $v$ with $\cub(v)\in [k_l,k_u]$, the whole partition needs to be loaded into main memory. When $k_u$ becomes small, it is highly possible for a partition to contain a node  $v$ with $\cub(v)\in [k_l,k_u]$. Consequently, almost all partitions are loaded into main memory. Due to this reason, the space used for \emcore is $O(m+n)$, and it cannot be significantly reduced in practice, as verified in our experiments. 

\stitle{Locality Property}. In this paper, we aim to design a semi-external algorithm for core decomposition. First, we introduce a locality property for core numbers, which is proposed in \cite{montresor2013}, as shown in the following theorem: 

\begin{theorem}
\label{thm:locality} \textbf{(Locality)}
Given a graph $G$, the $\core(v)$ values for all $v\in V(G)$ are their core numbers in $G$ iff:

\myitem There exists $V_k\subseteq \nb(v)$ such that:  $|V_k|=\core(v)$ and $\forall u\in V_k,\core(u)\geq \core(v)$; and\\

\vspace*{-0.5cm}
\myitem There does not exists $V_{k+1}\subseteq \nb(v)$ such that: $|V_{k+1}|=\core(v)+1$ and $\forall u\in V_{k+1}, \core(u) \geq \core(v)+1$.
\end{theorem}

Based on \refthm{locality}, the core number $\core(v)$ for a node $v\in V(G)$ can be calculated using the following recursive equation:

\vspace*{-0.7cm}
\begin{equation}
\label{eq:locality}
\core(v)= \text{max}\ k\ \text{s.t.}\ |\{u\in \nb(v)|\core(u)\geq k\}| \geq k
\end{equation}
\vspace*{-0.6cm}

Based on the locality property of core numbers, a distributed algorithm is designed in \cite{montresor2013}. In which each node $v$ initially assigns its core number as an arbitrary core number upper bound (e.g., $\deg(v)$), and keeps updating its core numbers using \refeq{locality} until convergence. 

\begin{algorithm}[t]
\caption{\small{$\semicore($Graph $G$ on Disk$)$}}
\label{alg:semicore}
\footnotesize
\begin{algorithmic}[1]
\STATE $\coreub(v)\leftarrow \dg(v)$ for all $v\in V(G)$;
\STATE $\update \leftarrow \textbf{true}$;

\WHILE{$\update$}
	\STATE $\update \leftarrow \textbf{false}$;
	\FOR{$v\leftarrow v_1$ \textbf{to} $v_n$}
		\STATE load $\nb(v)$ from disk;
		\STATE $\corg \leftarrow \coreub(v)$;
		\STATE $\coreub(v) \leftarrow \localcore(\corg, \nb(v))$;
		\STATE \textbf{if} $\coreub(v)\neq \corg$ \textbf{then} $\update \leftarrow \textbf{true}$;
	\ENDFOR
\ENDWHILE
\STATE \textbf{return} $\coreub(v)$ for all $v\in V(G)$;

\vspace*{0.2cm}

\STATE \textbf{Procedure} $\localcore(\corg$, $\nb(v))$
\STATE $\num(i)\leftarrow 0$ for all $1\leq i \leq c$;
\FORALL{$u\in \nb(v)$}
	\STATE $i\leftarrow \min\{\corg, \coreub(u)\}$;
	\STATE $\num(i)\leftarrow \num(i)+1$;
\ENDFOR
\STATE $s\leftarrow 0$;
\FOR{$k\leftarrow \corg$ \textbf{to} $1$}
	\STATE $s\leftarrow s + \num(k)$;
	\STATE \textbf{if} $s\geq i$ \textbf{then} \textbf{break};
\ENDFOR
\STATE \textbf{return} $k$;
\end{algorithmic}
\end{algorithm}

\stitle{Basic Solution}. In this paper, we make use of the locality property to design a semi-external algorithm for core decomposition. The pseudocode of our basic algorithm is shown in \refalg{semicore}. Here, we use $\coreub(v)$ to denote the intermediate core number for $v$, which is always an upper bound of $\core(v)$ and will finally converge to $\core(v)$. Initially, $\coreub(v)$ is assigned as an arbitrary upper bound of $\core(v)$ (e.g., $\deg(v)$). Then, we iteratively update $\coreub(v)$ for all $v\in V(G)$ using the locality property until convergence (line~2-9).

In each iteration (line~5-9), we sequentially scan the node table on disk to get the offset and $\dg(v)$ for each node $v$ from $v_1$ to $v_n$ (line~5). Then we load $\nb(v)$ from disk using the offset and $\dg(v)$ for each such node $v$, (line~6). Recall that the edge table  on disk stores $\nb(v)$ from $v_1$ to $v_n$ sequentially. Therefore, we can load $\nb(v)$ easily using sequential scan of the edge table on disk. In line~7-9, we record the original core number $\corg$ of $v$ (line~7); compute an updated core number of $v$ using \refeq{locality} by invoking $\localcore(\corg,\nb(v))$(line~8); and continue the iteration if $\coreub(v)$ is updated (line~9). Finally, when $\coreub(v)$ for all $v\in V(G)$ keeps unchanged, we return them as their core numbers (line~10). 

The procedure $\localcore(\corg,\nb(v))$ to compute the new core number of $v$ using \refeq{locality} is shown in line~11-20 of \refalg{semicore}. We use $\num(i)$ to denote the number of neighbors of $v$ with $\coreub$ equals $i$ (if $i<\corg$) or with $\coreub$ no smaller than $i$ (if $i=\corg$) (line~12-15). After computing $\num(i)$ for all $1\leq i \leq \corg$, we decrease $k$ from $\corg$ to $1$ (line~17), and for each $k$, we compute the number of neighbors of $v$ with $\coreub \geq k$, denotes as $s$ (line~18), i.e., $s=|\{u\in \nb(v)|\core(u)\geq k\}|$. Once $s\geq k$, we get the maximum $k$ with $|\{u\in \nb(v)|\core(u)\geq k\}| \geq k$, and we return $k$ as the new core number (line~20). Since $\corg\leq \nb(v)$, the time complexity of $\localcore(\corg,\nb(v))$ is $O(\dg(v))$. 

\stitle{Algorithm Analysis}. The space, CPU time, I/O complexities of \refalg{semicore} is shown in the following theorem:

\begin{theorem} \refalg{semicore} requires $O(n)$ memory.  Let $l$ be the number of iterations of \refalg{semicore}, the I/O complexity of \refalg{semicore} is $O(\frac{l\cdot(m+n)}{B})$, and the CPU time complexity of \refalg{semicore} is $O(l\cdot (m+n))$. 
\end{theorem}

\myproof First, three in-memory arrays are used in \refalg{semicore}, $\coreub$, $\num$, and $\nb$, all of which can be bounded using $O(n)$ memory. Consequently, \refalg{semicore} requires $O(n)$ memory. Second, in each iteration, \refalg{semicore} scans the node table and edge table sequentially once. Therefore, \refalg{semicore} consumes $O(\frac{l\cdot(m+n)}{B})$ I/Os. Finally, in each iteration, for each node $v$, we invoke $\localcore(\corg,\nb(v))$ which requires $O(\dg(v))$ CPU time. As a result, the CPU time complexity for \refalg{semicore} is $O(l\cdot (m+n))$. \eop

\stitle{Discussion}. Note that we use a value $l$ to denote the number of iterations of  \refalg{semicore}. Although $l$ is bounded by $n$ as proved in \cite{montresor2013}, it is much smaller in practice and is usually not largely influenced by the size of the graph. For example, in a social network \twitter with $n=41.7$ M, $m=1.47$ G, and $\kmax=2488$ used in our experiments, the number of iterations using \refalg{semicore} is only $62$. In a web graph \uk with $n=105.9$ M, $m=3.74$ G, and $\kmax=5704$ used in our experiments, the number of iterations is $2137$.  In the largest dataset \clueweb with $n=978.4$ M, $m=42.57$ G, and $\kmax=4244$ used in our experiments, the number of iterations is only $943$. 

\vspace*{-0.1cm}
\begin{figure}[h]
\begin{center}
{\small
\def\arraystretch{1.2}
\begin{tabular}{|l|l|l|l|l|l|l|l|l|l|} \hline
 \cp Iteration$\setminus$$v$&\cy$v_0$&\cy$v_1$&\cy$v_2$&\cy$v_3$&\cy$v_4$&\cy$v_5$&\cy$v_6$&\cy$v_7$&\cy$v_8$
 \\ \hline
 \cy Init   &  $3$ &  $3$ & $4$ & $6$ & $3$ &  $5$ & $3$ & $2$ & $1$ \\\hline\hline
 \cy Iteration $1$ & \cg $3$ & \cg $3$ & \cg $3$ & \cg $3$ & \cg $3$ & \cg $3$ & \cg $2$ & \cg $2$ & \cg $1$ \\\hline
 \cy Iteration $2$ & \cg $3$ & \cg $3$ & \cg $3$ & \cg $3$ & \cg $3$ & \cg $2$ & \cg $2$ & \cg $2$ & \cg $1$ \\\hline
 \cy Iteration $3$ & \cg $3$ & \cg $3$ & \cg $3$ & \cg $3$ & \cg $2$ & \cg $2$ & \cg $2$ & \cg $2$ & \cg $1$ \\\hline
 \cy Iteration $4$ & \cg $3$ & \cg $3$ & \cg $3$ & \cg $3$ & \cg $2$ & \cg $2$ & \cg $2$ & \cg $2$ & \cg $1$ \\
 \hline
\end{tabular}
}
\end{center}
\vspace*{-0.3cm}
\topcaption{\small{Illustration of $\semicore$}}
\vspace*{-0.2cm}
\label{fig:ex:semicore}
\end{figure}

\vspace*{-0.1cm}
\begin{example} 
\label{ex:semicore}
The process to compute the core numbers for nodes in \reffig{core} using \refalg{semicore} is shown in \reffig{ex:semicore}. The number in each cell is the value $\coreub(v_i)$ for the corresponding node $v_i$ in each iteration. The grey cells are those whose upper bounds is computed through invoking $\localcore$.  In iteration 1, when processing $v_3$, the $\coreub$ values for the neighbors of $v_3$ are $\{3, 3, 3, 3, 5, 3\}$. There are $3$ neighbors with $\coreub \geq 3$ but no $4$ neighbors with $\coreub \geq 4$. Therefore, $\coreub(v_3)$ is updated from $6$ to $3$. The algorithm terminates in $4$ iterations. 
\end{example}
\vspace*{-0.4cm}
\subsection{Partial Node Computation}
\label{sec:decomposition:partial}

\stitle{The Rationality}. In this subsection, we try to reduce the CPU and I/O consumption of \refalg{semicore}. Recall that in \refalg{semicore}, for all nodes $v\in V(G)$ in each iteration, the neighbors of $v$ are loaded from disk and $\coreub(v)$ is recomputed. However, if we can guarantee that $\coreub(v)$ is unchanged after recomputation, there is no need to load the neighbors of $v$ from disk and recompute $\coreub(v)$ by invoking $\localcore$. 

\vspace*{-0.3cm}
\begin{figure}[h]
\begin{center}
\begin{tabular}[t]{c}
\hspace*{-0.3cm}
\subfigure[\twitter (Vary Iteration)]{
	\includegraphics[width=0.5\columnwidth]{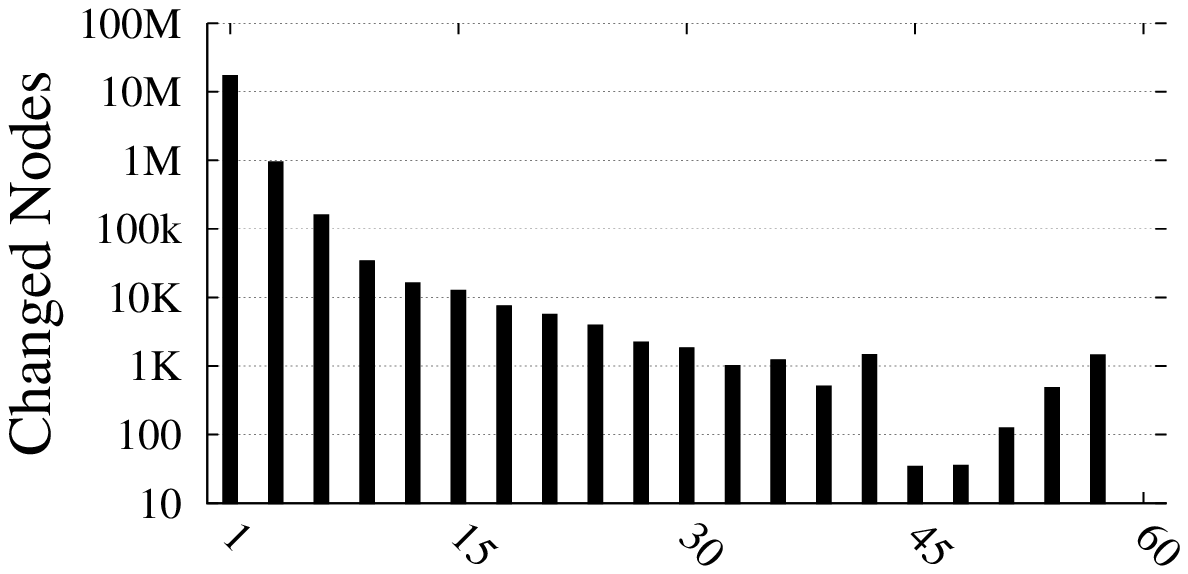}
}\hspace*{-0.3cm}
\subfigure[ \uk (Vary Iteration)]{
	\includegraphics[width=0.5\columnwidth]{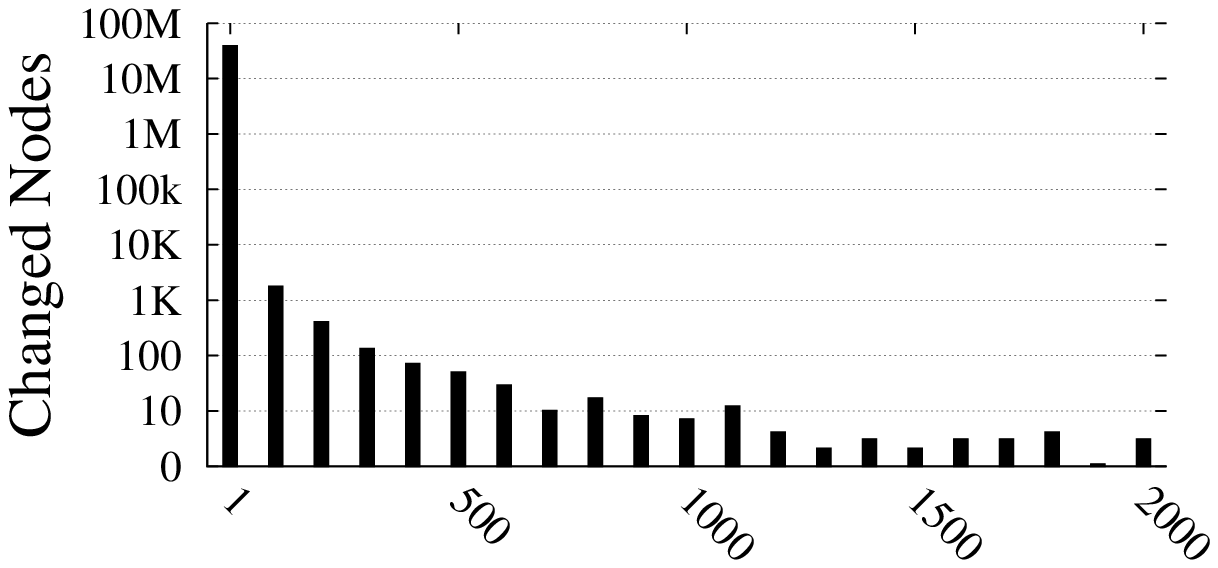}
}
\end{tabular}
\end{center}
\vspace*{-0.4cm}
\topcaption{{\small Number of Nodes Whose Core Numbers are Changes}}
\vspace*{-0.2cm}
\label{fig:rationality}
\end{figure}

To illustrate the effectiveness of eliminating such useless node computation, in \reffig{rationality},  we show the number of nodes whose $\coreub$ values are updated in each iteration for the \twitter and \uk datasets used in our experiments. In the \twitter dataset, totally $62$ iterations are involved. In iteration $1$, $10$ M nodes have their core numbers updated. However, in iteration $5$, only $1$ M nodes have their core numbers updated, which is only $10\%$ of the number in iteration $1$. From iteration $30$ on, less then $2$ K nodes have their core numbers updated in each iteration. In the \uk dataset, we have similar observation. There are totally $2137$ iterations. The number of core number updates in iteration $1$ is $10^4$ times larger than that in iteration $100$, and from iteration $400$ to iteration $2137$, less than $100$ nodes have their core numbers updated in each iteration.   

The above observations indicate that reducing the number of useless node computations can largely improve the performance of the algorithm. 

\stitle{Algorithm Design}. To reduce the useless node computations, we investigate a necessary condition for the core number of a node to be updated. According to \refeq{locality}, for a node $v$, if no core numbers of its neighbors are changed, the core number of $v$ will not change. Therefore, the following lemma can be easily derived.

\begin{lemma}
\label{lem:partial}
For each node $v\in V(G)$, $\coreub(v)$ is updated in iteration $i$ ($i>1$) only if there exists $u\in \nb(v)$ s.t. $\coreub(u)$ is updated in iteration $i-1$. 
\end{lemma}

Based on \reflem{partial}, in our algorithm, we use $\act(v)$ to denote whether node $v$ can be updated in each iteration. Only nodes $v$ with $\act(v)=$ true need to load their neighbors and have $\coreub(v)$ recomputed. The change of $\coreub(v)$ will trigger its neighbors $u\in \nb(v)$ to assign $\act(u)$ as true. We also maintain two values $v_{min}$ and $v_{max}$, which are the minimum node and maximum node with $\act=$ true. With $v_{min}$ and $v_{max}$, we can avoid checking all nodes in each iteration. Instead, we only need to check those nodes in the range from node $v_{min}$ to node $v_{max}$ for possible updates. 

\begin{algorithm}[t]
\caption{\small{$\semicorep($Graph $G$ on Disk$)$}}
\label{alg:semicorep}
\footnotesize
\begin{algorithmic}[1]
\STATE $\coreub(v)\leftarrow \dg(v)$ for all $v\in V(G)$;
\STATE $\act(v)\leftarrow \textbf{true}$ for all $v\in V(G)$;
\STATE $v_{min}\leftarrow v_1$; $v_{max}\leftarrow v_n$;
\STATE $\update \leftarrow \textbf{true}$;
\WHILE{$\update$}
	\STATE $\update \leftarrow \textbf{false}$; $v'_{min}\leftarrow v_n$; $v'_{max}\leftarrow v_1$;
	\FOR{$v\leftarrow v_{min}$ \textbf{to} $v_{max}$ s.t. $\act(v)=\textbf{true}$}
			\STATE $\act(v)\leftarrow \textbf{false}$;
			\STATE load $\nb(v)$ from disk;
			\STATE $\corg\leftarrow \coreub(v)$; $\coreub(v) \leftarrow \localcore(\corg, \nb(v))$;
			\IF{$\coreub(v)\neq c_{org}$} 
				\FORALL{$u\in \nb(v)$}
					\STATE $\act(u)\leftarrow \textbf{true}$;
					\STATE $\updaterange(v'_{min}, v'_{max}, v_{max}, \update, u, v )$;
					
				\ENDFOR
			\ENDIF
	\ENDFOR
	\STATE $v_{min}\leftarrow v'_{min}$; $v_{max}\leftarrow v'_{max}$;
\ENDWHILE
\STATE \textbf{return} $\coreub(v)$ for all $v\in V(G)$;

\vspace*{0.2cm}
\STATE \textbf{Procedure} $\updaterange(v'_{min}, v'_{max}, v_{max}, \update, u, v)$
\STATE $v_{max}\leftarrow \max\{v_{max}, u\}$
					\IF{$u<v$}
						\STATE $\update \leftarrow \textbf{true}$;
						\STATE $v'_{min} \leftarrow  \min\{v'_{min}, u\}$; $v'_{max}\leftarrow \max\{v'_{max}, u\}$;
					\ENDIF	
\end{algorithmic}
\end{algorithm}

Our algorithm $\semicorep$ is shown in \refalg{semicorep}. In line~1-4, we initialize $\coreub(v)$, $\act(v)$, $v_{min}$, $v_{max}$, and $\update$. We iteratively update the core numbers for nodes in $G$ until convergence. We use $v'_{min}$ and $v'_{max}$ to record the minimum and maximum nodes to be checked in the next iteration (line~6). In each iteration, we only check nodes from $v_{min}$ to $v_{max}$, and only recompute those nodes $v$ with $\act(v)$ being true (line~7). For each such $v$, we load $\nb(v)$ from disk and recompute its core number (line~8-10). If the core number of $v$ decreases, for each neighbor $u$ of $v$, we set $\act(u)$ to be true (line~11-13), and update $v'_{min}$, $v'_{max}$, $v_{max}$ by invoking $\updaterange(v'_{min}, v'_{max}, v_{max}, \update, u, v)$. The procedure $\updaterange$ is shown in line~17-21. We update $v_{max}$ using $u$ (line~18), since if $u>v$, $u$ can be computed in the current iteration other than be delayed to the next iteration. Only when $u<v$, we update the $v'_{min}$ and $v'_{max}$ for the next iteration using $u$ and set $\update$ to be true (line~19-21). After each iteration, we update $v_{min}$ and $v_{max}$ for next iteration (line~15). Finally, when the algorithm converges, we return $\coreub(v)$ for all $v\in V(G)$ as their core numbers (line~16). 

\vspace*{-0.1cm}
\begin{figure}[h]
\begin{center}
{\small
\def\arraystretch{1.2}
\begin{tabular}{|l|l|l|l|l|l|l|l|l|l|} \hline
 \cp Iteration$\setminus$$v$&\cy$v_0$&\cy$v_1$&\cy$v_2$&\cy$v_3$&\cy$v_4$&\cy$v_5$&\cy$v_6$&\cy$v_7$&\cy$v_8$
 \\ \hline
 \cy Init   & $3$ & $3$ & $4$ &  $6$ & $3$ & $5$ & $3$ & $2$ & $1$ \\\hline\hline
 \cy Iteration $1$ & \cg $3$ & \cg $3$ & \cg $3$ & \cg $3$ & \cg $3$ & \cg $3$ & \cg $2$ & \cg $2$ & \cg $1$ \\\hline
 \cy Iteration $2$ & \cg $3$ & \cg $3$ & \cg $3$ & \cg $3$ & \cg $3$ & \cg $2$ & \cg $2$ & \cg $2$ & \cg $1$ \\\hline
 \cy Iteration $3$ & $3$ & $3$ & $3$ & \cg $3$ & \cg $2$ & \cg $2$ & $2$ &  $2$ &  $1$ \\\hline
 \cy Iteration $4$ & $3$ & $3$ & \cg $3$ & \cg $3$ & $2$ & $2$ & $2$ & $2$ & $1$ \\
 \hline
\end{tabular}
}
\end{center}
\vspace*{-0.3cm}
\topcaption{\small{Illustration of $\semicorep$}}
\vspace*{-0.2cm}
\label{fig:ex:semicorep}
\end{figure}

\vspace*{-0.1cm}
\begin{example}
\label{ex:semicorep}
\reffig{ex:semicorep} shows the recomputed nodes (in grey cells) and their core numbers to process the graph shown in \reffig{core} using \refalg{semicorep}. In iteration $2$, the $\coreub(v_5)$ is updated from $3$ to $2$. This triggers its larger neighbors $v_6$, $v_7$, and $v_8$ to be computed in the same iteration, and its smaller neighbors $v_3$ and $v_4$ to be computed in the next iteration. Compared to \refalg{semicore} in \refex{semicore}, \refalg{semicorep} reduce the number of node computations from $36$ to $23$. 
\end{example}

\vspace*{-0.2cm}
\subsection{Optimal Node Computation}
\label{sec:decomposition:optimal}

Although $\semicorep$ improves $\semicore$ using partial node computation, it still involves a large number of useless node computations. For instance, in iteration $2$ of \refex{semicorep}, $\semicorep$ performs $9$ node computations, while only $1$ node updates its core number. In this section, we aim to design an optimal node computation scheme in the sense that every node computation will be guaranteed to update its core number. 

\stitle{The Rationality}.  Our general idea is to maintain more node information which can be used to check whether a node computation is needed.  Note that $\coreub(v)$ for each $v\in V(G)$ will not increase during the whole algorithm, and according to \refeq{locality}, $\core(v)$ is determined by the number of neighbors $u$ with $\core(u)\ge\core(v)$. Therefore, for each node $v$ in the graph, we maintain the number of such neighbors, denoted by $\cnt(v)$, which is defined as follows:

\vspace*{-0.5cm}
\begin{equation}
\label{eq:semicores}
\cnt(v)=|\{u\in \nb(v) | \coreub(u)\geq \coreub(v)\}|
\end{equation}
\vspace*{-0.6cm}

With the assistance of $\cnt(v)$ for all $v\in V(G)$, we can derive a sufficient and necessary condition for the core number of a node to be updated using the following lemma:

\begin{lemma}
\label{lem:semicores}
For each node $v\in V(G)$, $\coreub(v)$ is updated if and only if $\cnt(v)<\coreub(v)$.
\end{lemma}

\myproof We first prove $\Leftarrow$: Suppose $\cnt(v)<\coreub(v)$, we have $|\{u\in \nb(v) | \coreub(u)\geq \coreub(v)\}|<\coreub(v)$. Consequently, $\coreub(v)$ needs to be decreased by at least $1$ to satisfy \refeq{locality}. 

Next, we prove $\Rightarrow$: Suppose $\coreub(v)$ needs to be updated. According to \refeq{locality}, either $|\{u\in \nb(v) | \coreub(u)\geq \coreub(v)\}|<\coreub(v)$ or there is a larger $k$ s.t. $|\{u\in \nb(v) | \coreub(u)\geq k\}|\geq k$. The latter is impossible since $\coreub(u)$ will never increase during the algorithm. Therefore, $\cnt(v)<\coreub(v)$. \eop

\begin{algorithm}[t]
\caption{\small{$\semicores($Graph $G$ on Disk$)$}}
\label{alg:semicores}
\footnotesize
\begin{algorithmic}[1]
\STATE $\coreub(v)\leftarrow \dg(v)$ for all $v\in V(G)$;
\STATE $\cnt(v)\leftarrow 0$ for all $v\in V(G)$;
\STATE $v_{min}\leftarrow v_1$; $v_{max}\leftarrow v_n$;
\STATE $\update \leftarrow \textbf{true}$;
\WHILE{$\update$}
	\STATE $\update \leftarrow \textbf{false}$; $v'_{min}\leftarrow v_n$; $v'_{max}\leftarrow v_1$;
	\FOR{$v\leftarrow v_{min}$ \textbf{to} $v_{max}$ s.t. $\cnt(v)<\coreub(v)$}
			\STATE load $\nb(v)$ from disk;
			\STATE $\corg\leftarrow \coreub(v)$; $\coreub(v) \leftarrow \localcore(\corg, \nb(v))$;
			\STATE $\cnt(v)\leftarrow \computecnt(\nb(v), \coreub(v))$;
			\STATE $\updatenbrcnt(\nb(v), \corg, \coreub(v))$;
			\FORALL{$u\in \nb(v)$ s.t. $\cnt(u)<\coreub(u)$}
				\STATE $\updaterange(v'_{min}, v'_{max}, v_{max}, \update, u, v)$;
			\ENDFOR
	\ENDFOR
	\STATE $v_{min}\leftarrow v'_{min}$; $v_{max}\leftarrow v'_{max}$;
\ENDWHILE
\STATE \textbf{return} $\coreub(v)$ for all $v\in V(G)$;

\vspace*{0.2cm}
\STATE \textbf{Procedure} $\computecnt(\nb(v), \coreub(v))$
\STATE $s\leftarrow 0$;
\FORALL{$u\in \nb(v)$}
	\STATE \textbf{if} $\coreub(u) \geq \coreub(v)$ \textbf{then} $s\leftarrow s+1$;
\ENDFOR
\STATE \textbf{return} $s$;

\vspace*{0.2cm}
\STATE \textbf{Procedure} $\updatenbrcnt(\nb(v), \corg, \coreub(v))$
\FORALL{$u\in \nb(v)$}
	\IF{$\coreub(u)>\coreub(v)$ \textbf{and} $\coreub(u)\leq \corg$}
		\STATE $\cnt(u)\leftarrow \cnt(u)-1$;
	\ENDIF
\ENDFOR

\end{algorithmic}
\end{algorithm}

\stitle{Algorithm Design}. Based on the above discussion, we propose a new algorithm $\semicores$ with optimal node computation. The algorithm is shown in \refalg{semicores}. The initialization phase is similar to that in \refalg{semicorep} (line~1-4). For $\cnt(v)$ ($v\in V(G)$), we initialize it to be $0$ which will be updated to its real value after the first iteration. In each iteration, we partially scan the graph on disk similar to \refalg{semicorep} (line~6-14). Here, for each node $v$, the condition to load $\nb(v)$ from disk is $\cnt(v)<\coreub(v)$ according to \reflem{semicores} (line~7-8). In line~9, we compute the new $\coreub(v)$, and we can guarantee that $\coreub(v)$ will decrease by at least $1$. In line~10, we compute $\cnt(v)$ by invoking $\computecnt(\nb(v),\coreub(v))$  (line~16-20) which follows \refeq{semicores}. In line~11, since $\coreub(v)$ has been decreased from $\corg$, we need to update $\cnt(u)$ for every $u\in \nb(v)$ by invoking $\updatenbrcnt(\nb(v), \corg, \coreub(v))$ (line~21-24). Here, according to \refeq{semicores}, only those nodes $u$ with $\coreub(u)$ falling in the range $(\coreub(v),\corg]$ will have $\cnt(u)$ decreased by $1$ (line~23-24). In line~12-13, we need to update $v'_{min}$, $v'_{max}$, $v_{max}$, and $\update$ using those $u\in \nb(v)$ with $\cnt(u)<\coreub(u)$ (\reflem{semicores}). Here, we invoke the procedure $\updaterange$, which is the same as that used in \refalg{semicorep}. Finally, after the algorithm converges, the final core numbers for nodes in the graph are returned (line~15). 

Compared to \refalg{semicorep}, on the one hand, \refalg{semicores} can largely reduce the number of node computations since \refalg{semicores} only computes the core number of a node whenever necessary. On the other hand, for each node $v$ to be computed, in addition to invoking $\localcore$, \refalg{semicores} takes extra cost to maintain $\cnt(v)$ using $\computecnt$, and update $\cnt(u)$ for $u\in \nb(v)$ using $\updatenbrcnt$. However, it is easy to see that both $\computecnt$ and $\updatenbrcnt$ take $O(\dg(v))$ time which is the same as the time complexity of $\localcore$. Therefore, the extra cost can be well bounded. 

\stitle{Algorithm Analysis}. Compared to the state-of-the-art I/O efficient core decomposition algorithm $\emcore$, $\semicores$ (\refalg{semicores}) has the following advantages: 

\sstitle{$A_1$: Bounded Memory}. $\semicores$ follows the semi-external model and requires only $O(n)$ memory while $\emcore$ requires $O(m+n)$ memory in the worst case. For instance, to handle the \orkut dataset with $3$ M nodes and $117.2$ M edges used in our experiments, $\semicores$ consumes $12$ M memory; $\emcore$ consumes $938$ M memory; and the in-memory algorithm \imcore (\refalg{imcore}) consumes $1070$ M memory. 

\sstitle{$A_2$: Read I/O Only.} In $\semicores$, we only require read I/Os by scanning the node and edge tables sequentially on disk in each iteration. However, $\emcore$ needs both read and write I/Os since the partitions loaded into main memory will be repartitioned and written back to disk in each iteration. In practice, a write I/O is usually much slower than a read I/O. 

\sstitle{$A_3$: Simple In-memory Structure and Data Access. }  In $\emcore$, it invokes the in-memory algorithm $\imcore$ that uses a complex data structure for bin sort. It also involves complex graph partitioning and repartitioning algorithms. In $\semicores$, we only use two arrays $\core$ and $\cnt$, and the data access is simple. This makes $\semicores$ very efficient in practice and even more efficient than the in-memory algorithm $\imcore$ in many datasets. For instance, to handle the \orkut dataset used in our experiments, $\emcore$, $\imcore$, and $\semicores$ consumes $63.2$ seconds, $18.4$ seconds, and $16.3$ seconds respectively. 

\vspace*{-0.1cm}
\begin{figure}[h]
\begin{center}
{\small
\def\arraystretch{1.2}
\begin{tabular}{|l|l|l|l|l|l|l|l|l|l|} \hline
 \cp Iteration$\setminus$$v$&\cy$v_0$&\cy$v_1$&\cy$v_2$&\cy$v_3$&\cy$v_4$&\cy$v_5$&\cy$v_6$&\cy$v_7$&\cy$v_8$
 \\ \hline
 \cy Init   & $3$ & $3$ & $4$ & $6$ & $3$ & $5$ & $3$ & $2$ & $1$ \\\hline \hline
 \cy Iteration $1$ & \cg $3$ & \cg $3$ & \cg $3$ & \cg $3$ & \cg $3$ & \cg $3$ & \cg $2$ & \cg $2$ & \cg $1$ \\\hline
 \cy Iteration $2$ & $3$ & $3$ & $3$ & $3$ & $3$ & \cg $2$ & $2$ & $2$ & $1$ \\\hline
 \cy Iteration $3$ & $3$ & $3$ & $3$ & $3$ & \cg $2$ &  $2$ & $2$ &  $2$ &  $1$ \\\hline
\end{tabular}
}
\end{center}
\vspace*{-0.3cm}
\topcaption{\small{Illustration of $\semicores$}}
\vspace*{-0.2cm}
\label{fig:ex:semicores}
\end{figure}

\vspace*{-0.1cm}
\begin{example}
\label{ex:semicores}
The process to handle the graph $G$ in \reffig{core} using \refalg{semicores} is shown in \reffig{ex:semicores}. We show $\coreub(v)$ each $v\in V(G)$ in each iteration, and those recomputed $\coreub(v)$ values are shown in grey cells.  For instance, after iteration $1$, we have $\coreub(v_5)=3$ and $\cnt(v_5)=2$ since only its two neighbors $v_3$ and $v_4$ have their $\coreub$ values no smaller than $3$. Therefore, in iteration $2$, $\coreub(v_5)$ is recomputed and updated from $3$ to $2$. This also updates the $\cnt$ value of its neighbor $v_4$ from $3$ to $2$ since $\coreub(v_4)=3$. Note that in iteration $1$, we need to compute $\coreub(v)$ for all $v\in V(G)$ since $\cnt(v)$ is unknown only in the first iteration. Compared to \refalg{semicorep} in \refex{semicorep}, \refalg{semicores} only uses $3$ iterations and reduces the number of node computations from $23$ to $11$. 
\end{example}

\vspace*{-0.2cm}
\section{I/O Efficient Core Maintenance}
\label{sec:maintenance}
In this section, we discuss how to incrementally maintain the core numbers when edges are inserted  into or deleted from the graph under the semi-external setting. 

\subsection{Edge Deletion}
\label{sec:maintenance:deletion}

\stitle{Algorithm Design}. In \refthm{change1}, we know that after an edge deletion, the core number for any $v\in V(G)$ will decrease by at most $1$. Therefore, after an edge deletion, the old core numbers of nodes in the graph are upper bounds of their new core numbers.  Recall that in \refalg{semicores}, as long as $\coreub(v)$ is initialized to be an arbitrary upper bound of $\core(v)$ for all $v\in V(G)$, $\coreub(v)$ can be finally converged to $\core(v)$ after the algorithm terminates. Therefore, \refalg{semicores} can be easily modified to handle edge deletion. 

\begin{algorithm}[t]
\caption{\small{$\semidels($Graph $G$ on Disk, Edge $(u,v))$}}
\label{alg:semidels}
\footnotesize
\begin{algorithmic}[1]
\STATE delete $(u,v)$ from $G$;
\IF{$\coreub(u)<\coreub(v)$}
	\STATE $\cnt(u)\leftarrow \cnt(u)-1$;
	\STATE $v_{min}\leftarrow u$; $v_{max}\leftarrow u$;
\ELSIF{$\coreub(v)<\coreub(u)$}
	\STATE $\cnt(v)\leftarrow \cnt(v)-1$;
	\STATE $v_{min}\leftarrow v$; $v_{max}\leftarrow v$;
\ELSE
	\STATE $\cnt(u)\leftarrow \cnt(u)-1$; $\cnt(v)\leftarrow \cnt(v)-1$;
	\STATE $v_{min}\leftarrow \min\{u,v\}$; $v_{max}\leftarrow \max\{u,v\}$;
\ENDIF
\STATE line~4-14 of \refalg{semicores};
\end{algorithmic}
\end{algorithm}

Specifically, we show our algorithm $\semidels$ for edge deletion in \refalg{semidels}. Given an edge $(u,v)\in E(G)$ to be removed, we first delete $(u,v)$ from $G$ (line~1). We will discuss how to update $G$ on disk after edge deletion / insertion later.  In line~2-8, we update $\cnt(u)$ and $\cnt(v)$ due to the deletion of edge $(u,v)$, and we also compute the initial range $v_{min}$ and $v_{max}$ for node checking. Here, we consider three cases. First, if $\coreub(u)<\coreub(v)$, we only need to decrease $\cnt(u)$ by $1$, and set $v_{min}$ and $v_{max}$ to be $u$. Second, if $\coreub(v)<\coreub(u)$, we decrease $\cnt(v)$ by $1$, and set $v_{min}$ and $v_{max}$ to be $v$. Third, if $\coreub(v)=\coreub(u)$, we decrease both $\cnt(v)$ and $\cnt(u)$ by $1$, and set $v_{min}$ and $v_{max}$ to be $\min\{u,v\}$ and $\max\{u,v\}$ respectively. Now we can use \refalg{semicores} to update the core numbers of other nodes (line~11). 

\stitle{Graph Maintenance}. We introduce how to maintain the graph on disk when edges are inserted into / deleted from the graph. Recall that our graph is stored in terms of adjacency lists on disk. If we simply update the lists after each edge insertion / deletion, the cost will be too high. To handle this, we allow a memory buffer to maintain the latest inserted / deleted edges. We also index the edges in the memory buffer. When the buffer is full, we update the graph on disk and clear the buffer. Noticed that each time when we load $\nb(v)$ for a certain node $v$ from disk, we also need to obtain the inserted / deleted edges for $v$ from the memory buffer, and use them to compute the updated $\nb(v)$.  

\begin{figure}[h]
\begin{center}
{\small
\def\arraystretch{1.2}
\begin{tabular}{|l|l|l|l|l|l|l|l|l|l|} \hline
 \cp Iteration$\setminus$$v$&\cy$v_0$&\cy$v_1$&\cy$v_2$&\cy$v_3$&\cy$v_4$&\cy$v_5$&\cy$v_6$&\cy$v_7$&\cy$v_8$
 \\ \hline 
 \cy Old Value   & $3$ & $3$ & $3$ & $3$ & $2$ &  $2$ & $2$ &  $2$ &  $1$ \\\hline \hline
 \cy Iteration $1$ &  \cg $2$ & \cg $2$ & \cg $2$ & \cg $2$ & $2$ & $2$ & $2$ & $2$ & $1$ \\\hline
\end{tabular}
}
\end{center}
\vspace*{-0.3cm}
\topcaption{\small{Illustration of $\semidels$} (Delete $(v_0,v_1)$)}
\vspace*{-0.1cm}
\label{fig:ex:semidels}
\end{figure}

\begin{example} 
\label{ex:semidels} 
Suppose after \refex{semicores}, we delete edge $(v_0,v_1)$ from $G$ (\reffig{core}). Using \refalg{semidels}, we first update both $\cnt(v_0)$ and $\cnt(v_1)$ from $3$ to $2$ and then invoke line~4-14 of \refalg{semicores} with $v_{min}=0$ and $v_{max}=1$. Only $1$ iteration is needed with $4$ node computations as shown in \reffig{ex:semidels}.
\end{example}

\vspace*{-0.2cm}
\subsection{Edge Insertion}
\label{sec:maintenance:insertion}

\stitle{The Rationality}. After  a new edge $(u,v)$ is inserted into graph $G$, according to \refthm{change1}, we know that the core number for any $v\in V(G)$ will increase by at most $1$. As a result, the old core number of a node in the graph may not be an upper bound of its new core number. Therefore, \refalg{semicores} cannot be applied directly to handle edge insertion. However, according to \refthm{change2}, after inserting an edge $(u,v)$ (suppose $\coreub(v)\leq \coreub(u)$), we can find a candidate set $V_c$ consisting of all nodes $w$ that are reachable from node $v$ via a path that consists of nodes with $\coreub$ equals $\coreub(v)$, and we can guarantee that those nodes with core numbers increased by $1$ is a subset of $V_c$. Consequently, if we increase $\coreub(v)$ by $1$ for all $v\in V_c$, we can guarantee that for all $u\in V(G)$, $\coreub(u)$  is an upper bound of the new core number of $u$. Thus we can apply \refalg{semicores} to compute the new core numbers. 

\stitle{Algorithm Design}. Our algorithm $\semiadd$ for edge insertion is shown in \refalg{semiadd}. In line~1, we insert $(u,v)$ into $G$. In line~1-4, we update $\cnt(u)$ and $\cnt(v)$ caused by the insertion of edge $(u,v)$. We use $\act(w)$ to denote whether $w$ is a candidate node with core number increased which is initialized to be false except for node $u$. In line~8-21, we iteratively update $\act(w)$ for $w\in V(G)$ until convergency.  In each iteration (line~9-20), we find nodes  $v'$ with $\act(v')=$ true and $\coreub(v')$ not being increased (line~11). For each such node $v'$, we increase $\coreub(v')$ by $1$ (line~12), and load $\nb(v')$ from disk. Since $\coreub(v')$ is changed, we need to compute $\cnt(v')$ (line~14) and update the $\cnt$ values for the neighbors of $v'$ (line~15-16). In line~17-20, we set $\act(u')$ to be true for all the neighbors $u'$ of $v'$ (line~17-18) if $u'$ is a possible candidate (line~18), and we update the range of nodes to be checked in the next iteration (line~20). After all iterations, we compute the range of the candidate nodes (line~22-24). Now we can guarantee that $\coreub(v')$ is an upper bound of the new core number of $v'$. Therefore, we invoke line~4-14 of \refalg{semicores} to compute the core numbers of all nodes in the graph (line~25).

\begin{figure}[h]
\begin{center}
{\small
\def\arraystretch{1.2}
\begin{tabular}{|l|l|l|l|l|l|l|l|l|l|} \hline
 \cp Iteration$\setminus$$v$&\cy$v_0$&\cy$v_1$&\cy$v_2$&\cy$v_3$&\cy$v_4$&\cy$v_5$&\cy$v_6$&\cy$v_7$&\cy$v_8$
 \\ \hline
 \cy Old Value   & $2$ & $2$ & $2$ & $2$ & $2$ &  $2$ & $2$ &  $2$ &  $1$ \\\hline\hline
 \cy Iteration $1.1$ &  $2$ & $2$ & $2$ &  $2$ & \cg $3$ & \cg $3$ & \cg $3$ & \cg $3$ & $1$ \\\hline
 \cy Iteration $1.2$ & $2$ & $2$ & \cg $3$ & \cg $3$ &  $3$ &  $3$ &  $3$ &  $3$ & $1$ \\\hline
 \cy Iteration $1.3$ & \cg $3$ & \cg $3$ &  $3$ & $3$ &  $3$ &  $3$ &  $3$ &  $3$ & $1$ \\\hline \hline
 \cy Iteration $2.1$ & \cg $2$ & \cg $2$ & \cg $2$ & $3$ &  $3$ &  $3$ &  $3$ &  \cg $2$ & $1$ \\\hline
\end{tabular}
}
\end{center}
\vspace*{-0.3cm}
\topcaption{\small{Illustration of $\semiadd$} (Insert $(v_4,v_6)$)}
\vspace*{-0.2cm}
\label{fig:ex:semiadd}
\end{figure}

\begin{example}
\label{ex:semiadd}
Suppose after deleting edge $(v_0, v_1)$ from the graph $G$ (\reffig{core}) in \refex{semidels}, we insert a new edge $(v_4, v_6)$ into $G$.  The process to compute the new core numbers of nodes in $G$ is shown in \reffig{ex:semiadd}. Here, we use $3$ iterations $1.1$, $1.2$, and $1.3$ to compute the candidate nodes, and use $1$ iteration $2.1$ to compute the new core numbers. In iteration $1.1$, when $v_4$ is computed, it triggers its smaller neighbors $v_2$ and $v_3$ to be computed in the next iteration and triggers its larger neighbor $v_5$ to be computed in the current iteration. The total number of node computations is $12$. 
\end{example}

\begin{algorithm}[t]
\caption{\small{$\semiadd($Graph $G$ on Disk, Edge $(u,v))$}}
\label{alg:semiadd}
\footnotesize
\begin{algorithmic}[1]
\STATE insert $(u,v)$ into $G$;
\STATE swap $u$ and $v$ if $\coreub(u)>\coreub(v)$;
\STATE $\cnt(u)\leftarrow \cnt(u)+1$;
\STATE \textbf{if} $\coreub(v)=\coreub(u)$ \textbf{then} $\cnt(v)\leftarrow \cnt(v)+1$;
\STATE $\corg\leftarrow \coreub(u)$;
\STATE $\act(w)\leftarrow \textbf{false}$ for all $w\in V(G)$; $\act(u)\leftarrow$ \textbf{true}; 
\STATE $v_{min}\leftarrow u$; $v_{max}\leftarrow u$; $\update\leftarrow \textbf{true}$;
\WHILE{$\update$}
	\STATE $\update \leftarrow \textbf{false}$; $v'_{min}\leftarrow v_n$; $v'_{max}\leftarrow v_1$;
	\FOR{$v'\leftarrow v_{min}$ \textbf{to} $v_{max}$}
		\IF{$\act(v')=\textbf{true}$ \textbf{and} $\coreub(v')=\corg$}
			\STATE $\coreub(v')\leftarrow \coreub(v')+1$;
			\STATE load $\nb(v')$ from disk;
			\STATE $\cnt(v')\leftarrow \computecnt(\nb(v'), \coreub(v'))$;
			\FORALL{$u'\in \nb(v')$ s.t. $\coreub(u')=\coreub(v')$}
				\STATE $\cnt(u')\leftarrow \cnt(u')+1$;
			\ENDFOR
			\FORALL{$u'\in \nb(v')$}
				\IF{$\coreub(u')=\corg$ and $\act(u')=\textbf{false}$}
					\STATE $\act(u')\leftarrow \textbf{true}$; 
					\STATE $\updaterange(v'_{min}, v'_{max}, v_{max}, \update, u', v')$;
				\ENDIF
			\ENDFOR
		\ENDIF
	\ENDFOR
	\STATE $v_{min}\leftarrow v'_{min}$; $v_{max}\leftarrow v'_{max}$;
\ENDWHILE

\STATE $v_{min}\leftarrow u$; $v_{max}\leftarrow u$;
\FORALL{$v\in V(G)$ s.t. $\act(v)=\textbf{true}$}
	\STATE $v_{min}\leftarrow \min\{v_{min}, v\}$; $v_{max}\leftarrow \max\{v_{max}, v\}$;
\ENDFOR

\STATE line~4-14 of \refalg{semicores};
\end{algorithmic}
\end{algorithm}

\vspace*{-0.2cm}
\subsection{Optimization for Edge Insertion}
\label{sec:maintenance:insertionnew}

\stitle{The Rationality}. \refalg{semiadd} handles an edge insertion using two phases. In phase 1, we compute a superset $V_c$ of nodes whose core numbers will be updated, and we increase the core numbers for all nodes in $V_c$ by $1$. In phase 2, we compute the core numbers of all nodes using \refalg{semicores}. One problem of \refalg{semiadd} is that the size of $V_c$ can be very large, which may result in a large number of node computations and I/Os in both phase 1 and phase 2 of  \refalg{semiadd}. Therefore, it is crucial to reduce the size of $V_c$. 

Now, suppose and edge $(u,v)$ is inserted into the graph $G$; $\cnt(u)$ and $\cnt(v)$ are updated accordingly; and $\coreub(w)$ values for all $w\in V(G)$ have not been updated . Without loss of generality, we assume that $\coreub(u)<\coreub(v)$ and let $\corg=\coreub(u)$. Let $V_c$ be the set of candidate nodes computed in \refalg{semiadd}, i.e., $V_c$ consists of all nodes that are reachable from $u$ via a path that consists of nodes with $\coreub$ equals $\corg$. Let $V_c^*\subseteq V_c$ be the set of nodes with $\coreub$ updated to be $\corg+1$ after inserting $(u,v)$. We have the following lemmas:

\begin{lemma} 
\label{lem:o1}
(a) For $v'\in V_c\setminus V_c^*$, $\cnt(v')$ keeps unchanged; (b) For  $v'\in V_c^*$, $\cnt(v')$ will not increase.
\end{lemma}

\myproof This lemma can be easily verified according to \refeq{semicores} and \refthm{change1}. \eop

\begin{lemma}
\label{lem:o2}
 If $\cnt(v')\geq \corg+1$ for all $v'\in V_c$, then we have $V_c^*=V_c$.
\end{lemma}

\myproof If we increase $\coreub(v')$ by $1$ for all $v'\in V_c$, it is easy to verify that $\cnt(v')$ for all $v'\in V_c$ keep unchanged. Now suppose $\cnt(v')\geq \corg+1$ for all $v'\in V_c$, we can derive that the locality property in \refthm{locality} holds for every $v'\in V(G)$. Therefore, the new $\coreub(v')$ is the core number of $v'$ for every $v'\in V(G)$. This indicates that $V_c^*=V_c$. \eop

\begin{lemma}
\label{lem:o3}
For any $v'\in V_c$, if $v'\in V_c^*$, then we have $\cnt(v')\geq \corg+1$.
\end{lemma}

\myproof Since $v'\in V_c^*$, we know that the new $\cnt(v')$ is no smaller than $\corg+1$. According to \reflem{o1} (b), the original $\cnt(v')$ is also no smaller than $\corg+1$, since $\cnt(v')$ will not increase. Therefore, the lemma holds. \eop


\begin{theorem}
\label{thm:o4} For each $v'\in V_c$, we define $\cnts(v')$ as:
\\\vspace*{0.1cm}
\vspace*{-0.3cm}
\begin{equation}
\label{eq:o4}
\cnts(v') = |\{u'\in \nb(v')\ |\ \coreub(u')>\corg \ or \ u'\in V_c^*\}|
\end{equation}
\vspace*{-0.6cm}
\\\vspace*{0.1cm}
\noindent We have: 
\\\vspace*{0.1cm}
\noindent (a) If $v'\in V_c^*$, then the updated $\cnt(v')=\cnts(v')$; and
\\\vspace*{0.1cm}
\noindent (b) $v'\in V_c^* \Leftrightarrow \cnts(v')\geq \corg+1$.
\end{theorem} 

\myproof For (a): for all $v'\in V_c^*$, since $\coreub(v')$ will become $\corg+1$, all nodes  $u'\in V_c\setminus V_c^*$ will not contribute to $\cnt(v')$ according to  \refeq{semicores}. Therefore, (a) holds. 

\noindent For (b): $\Rightarrow$ can be derived according to (a). Now we prove $\Leftarrow$. Suppose $\cnts(v')\geq \corg+1$, to prove $v'\in V_c^*$, we prove that if we increase $\coreub(u')$ to $\corg+1$ for all $u'\in V_c$ and apply \refalg{semicores}, then $\coreub(v')$ will keep to be $\corg+1$ after convergency. Note that for all nodes $u' \in V_c^*$ and $u'\in \nb(v')$, $\coreub(u')$ will keep to be $\corg+1$ and will  contribute to $\cnt(v')$, and all nodes $u'\in \nb(v')$ with $\coreub(u')>\corg$ will also contribute to $\cnt(v')$. According to \refeq{o4}, we have $\cnt(v')\geq \cnts(v') \geq \corg+1$. Therefore, $\coreub(v')$ will never decrease according to \reflem{semicores}. This indicates that $v'\in V_c^*$. \eop

According to \refthm{o4} (b), $\cnts(v')$ can be defined using the following recursive equation:

\vspace*{-0.6cm}
\begin{multline}
\label{eq:cnts}
\cnts(v')=  |\{u'\in \nb(v')\ |\ \coreub(u')>\corg\ \text{or}\ \\(\coreub(u')=\corg\  \text{and}\ \cnts(u')\geq \corg+1)\}|
\end{multline}
\vspace*{-0.6cm}

To compute $\cnts(v')$ for all $v'\in V_c$, we can initialize $\cnts(v')$ to be $\cnt(v')$, and apply \refeq{cnts} iteratively on all $v'\in V_c$ until convergency.  However, this algorithm needs to compute $V_c$ first, which is inefficient. Note that according to \refeq{cnts} and \refthm{o4} (b), we only care about those nodes $u'$ with $\cnts(u')\geq \corg+1$. Therefore, we do not need to compute the whole $V_c$ by expanding from node $u$. Instead, for each expanded node $u'$, if we guarantee that $\cnts(u')< \corg+1$, we do not need to expand $u'$ further. In this way, the computational and I/O cost can be largely reduced. 

\begin{algorithm}[t]
\caption{\small{$\semiadds($Graph $G$ on Disk, Edge $(u,v))$}}
\label{alg:semiadds}
\footnotesize
\begin{algorithmic}[1]
\STATE line~1-5 of \refalg{semiadd};
\STATE $\status(w)\leftarrow \statuse$ for all $w\in V(G)$; $\status(u)\leftarrow \statusq$; 
\STATE $v_{min}\leftarrow u$; $v_{max}\leftarrow u$; $\update\leftarrow \textbf{true}$;
\WHILE{$\update$}
	\STATE $\update \leftarrow \textbf{false}$; $v'_{min}\leftarrow v_n$; $v'_{max}\leftarrow v_1$;
	\FOR{$v'\leftarrow v_{min}$ \textbf{to} $v_{max}$}
		\IF{$\status(v')=\statusq$}
			\STATE load $\nb(v')$ from disk;
			\STATE $\cnt(v')\leftarrow \computecnts(\nb(v'),\corg)$;
			\STATE $\status(v')\leftarrow \statusy$; $\coreub(v')\leftarrow \corg+1$;
			
			\FORALL {$u'\in \nb(v')$ s.t. $\coreub(u')=\corg+1$}
				\STATE $\cnt(u')\leftarrow \cnt(u')+1$;
			\ENDFOR
			
			\IF{$\cnt(v')\geq \corg+1$}
				\FORALL{$u'\in \nb(v')$ s.t. $\coreub(u')=\corg$}
					\IF{$\cnt(u') \geq \corg +1$ and $\status(u') = \statuse$}
						\STATE $\status(u')\leftarrow \statusq$; 
						\STATE $\updaterange(v'_{min}, v'_{max}, v_{max}, \update, u', v')$;
					\ENDIF
				\ENDFOR
			\ENDIF
		\ENDIF
		\IF{ $\status(v')=\statusy$ \textbf{and} $\cnt(v')< \corg+1$}
			\STATE load $\nb(v')$ from disk if not loaded;
			\STATE $\cnt(v')\leftarrow \computecnt(\nb(v'),\corg)$;
			\STATE $\status(v')\leftarrow \statusn$; $\coreub(v')\leftarrow \corg$;
			\FORALL {$u'\in \nb(v')$ s.t. $\coreub(u')=\corg+1$}
				\STATE $\cnt(u')\leftarrow \cnt(u')-1$;
			\ENDFOR
			\FORALL{$u'\in \nb(v')$ s.t. $\status(u')=\statusy$}
				\STATE $\cnt(u')\leftarrow \cnt(u')-1$; 
				\IF{$\cnt(u')<\corg+1$}
					\STATE $\updaterange(v'_{min}, v'_{max}, v_{max}, \update, u', v')$;
				\ENDIF
			\ENDFOR
		\ENDIF
	\ENDFOR
	\STATE $v_{min}\leftarrow v'_{min}$; $v_{max}\leftarrow v'_{max}$;
\ENDWHILE


\vspace*{0.2cm}
\STATE \textbf{Procedure} $\computecnts(\nb(v'),\corg)$
\STATE $s\leftarrow 0$;
\FORALL{$u'\in \nb(v')$}
	\STATE \hspace*{-0.2cm} \textbf{if} $\coreub(u')>\corg$ \textbf{or} ($\coreub(u')=\corg$ \textbf{and} $\cnt(u')\geq \corg+1$ \textbf{and} $\status(u')\neq \statusn$) \textbf{then} $s\leftarrow s+1$;
\ENDFOR
\STATE \textbf{return} $s$;
\end{algorithmic}
\end{algorithm}

\stitle{Algorithm Design}. Based on the above discussion, for each node $w\in V(G)$, we use $\status(w)$ to denote the status of node $w$ during the processing of node expansion. Each node $w\in V(G)$ has the following four status ($\status(w)$):  
\vspace*{0.2cm}
\\\vspace*{0.1cm}
\sstitle{$\statuse$}:  $w$ has not been expanded by other nodes.
\\\vspace*{0.1cm}
\sstitle{$\statusq$}:  $w$ is expanded but $\cnts(w)$ is not calculated.
\\\vspace*{0.1cm}
\sstitle{$\statusy$}: $\cnts(w)$ is calculated with $\cnts(w)\geq \corg+1$.
\\\vspace*{0.1cm}
\sstitle{$\statusn$}: $\cnts(w)$ is calculated with $\cnts(w)< \corg+1$.

With $\status(w)$ and according to \refthm{o4} (a) and \reflem{o1} (a), we can reuse $\cnt(w)$ to represent $\cnts(w)$ for each $w\in V(G)$. That is, if $\status(w)=\statusy$, $\cnt(w)$ can represent $\cnts(w)$ which is calculated using \refeq{cnts}, otherwise, if $\status(w)=\status$ $\cnt(w)$ is calculated using \refeq{semicores}. 

Our new algorithm $\semiadds$ for edge insertion is shown in \refalg{semiadds}. The initialization phase is similar to that in \refalg{semiadd} (line~1). In line~6, we initialize $\status(w)$ to be $\statuse$ except $\status(u)$ which is initialized to be $\statusq$. The algorithm iteratively update $\status(v')$, $\coreub(v')$, and $\cnt(v')$ for all $v'\in V(G)$. In each iteration (line~5-28), we check $v'$ from $v_{min}$ to $v_{max}$  (line~6), and for each such $v'$ to be checked, we consider the following status transitions: 

\sstitle{\myitem From $\statusq$ to $\statusy$ (line~7-12):} If $\status(v')=\statusq$ (line~7), we load $\nb(v')$ from disk (line~8) and compute $\cnt(v')$ using \refeq{cnts} by invoking $\computecnts(\nb(v'),\corg)$ which is shown in line~29-33. Compared to \refeq{cnts}, we add a new condition for $u'\in \nb(v')$: $\status(u')\neq \statusn$ (line~32). This is because for node $u'$ with $\status(u')= \statusn$, it is computed using \refeq{semicores} other than \refeq{cnts}, and it cannot contribute to $\cnt(v')$. After computing $\cnt(v')$, in line~10, we set $\status(v')$ to be $\statusy$ and increase $\coreub(v')$ to be $\corg+1$. Since $\coreub(v')$ is increased to be $\corg+1$, we need to increase $\cnt(u')$ for all neighbor $u'$ of $v'$ with $\coreub(u')=\corg+1$ (line~11-12). 

\sstitle{\myitem From $\statuse$ to $\statusq$ (line~13-17):} After setting $v'$ to be $\statusq$, if $\cnt(v')\geq \corg+1$, $v'$ will not set to be $\statusn$ in this iteration. In this case (line~13),  we can expand $v'$. That is, for all neighbors $u'$ of $v'$ with $\coreub(u')=\corg$ (line~14), if $\cnt(u')\geq \corg+1$ (refer to \reflem{o3}) and $u'$ has not be expanded ($\status(u')=\statuse$), we set $\status(u')$ to be $\statusq$ so that $u'$ can be expanded, and update the range of nodes to be checked (line~15-17). 

\sstitle{\myitem From $\statusy$ to $\statusn$ (line~18-27):} If $\status(v')$ is $\statusy$ and $\cnt(v')<\corg+1$, we need to change the status of $v'$ (line~18). Here, in line~19, we load $\nb(v')$ from disk if it is not loaded in line~8. In line~20, we compute $\cnt(v')$ using \refeq{semicores}. In line~21, we set $\status(v')$ to be $\statusn$, and update $\coreub(v')$ to be $\corg$ according to \reflem{o1} (a). Since $\coreub(v')$ is changed from $\corg+1$ to $\corg$, for all neighbors $u'$ of $v'$ with $\coreub(u')=\corg+1$, we need to decrease $\cnt(u')$ (line~22-23). In addition, according to \refeq{cnts}, the status change from $\statusy$ to $\statusn$ for $v'$ will trigger each neighbor $u'$ of $v'$ to decrease its $\cnt(u')$ if $\status(u')=\statusy$ (line~24-25). For each such $u'$, if $\cnt(u')$ is decreased below $\corg$, $\status(u')$ need to be updated in the same of later iterations (line~26-27). 

Compared to \refalg{semiadd} that requires two phases to update the core numbers, \refalg{semiadds} requires only one phase without invoking \refalg{semicores} for core number updates. 

\begin{figure}[h]
\begin{center}
{\small
\def\arraystretch{1.2}
\begin{tabular}{|l|l|l|l|l|l|l|l|l|l|} \hline
 \cp Iteration$\setminus$$v$&\cy$v_0$&\cy$v_1$&\cy$v_2$&\cy$v_3$&\cy$v_4$&\cy$v_5$&\cy$v_6$&\cy$v_7$&\cy$v_8$
 \\ \hline
 \cy Old Value   & $2$ & $2$ & $2$ & $2$ & $2$ &  $2$ & $2$ &  $2$ &  $1$ \\\hline \hline
 \cy Iteration $1$ &  \statuse & \statuse & \statusq & \statusq & \cg \statusy & \cg \statusy & \cg \statusy & \statuse & \statuse \\\hline
 \cy Iteration $2$ & \statuse  & \statuse & \cg \statusn & \cg \statusy &  \statusy &  \statusy &  \statusy &  \statuse & \statuse \\\hline \hline
  \cy New Value & $2$ & $2$ & $2$ & $3$ &  $3$ &  $3$ &  $3$ &  $2$ & $1$ \\\hline
\end{tabular}
}
\end{center}
\vspace*{-0.3cm}
\topcaption{\small{Illustration of $\semiadds$} (Insert $(v_4,v_6)$)}
\vspace*{-0.2cm}
\label{fig:ex:semiadds}
\end{figure}

\begin{example}
\label{ex:semiadds}
Suppose after deleting edge $(v_0, v_1)$ from graph $G$ (\reffig{core}) in \refex{semidels}, we insert edge $(v_4, v_6)$ into $G$. The process to update the status of nodes in each iteration is shown in \reffig{ex:semiadds}. In iteration $1$, when we check $v_4$, we update $\status(v_4)$ from $\statusq$ to be $\statusy$, and update the status of its neighbors ($v_2$, $v_3$, $v_5$, and $v_6$) to be $\statusq$. In iteration $2$, for $v_2$ with status $\statusq$, we can calculate that $\cnt(v_2)=2<\corg+1=3$. Therefore, we set $\status(v_2)$ to be $\statusn$, and decrease $\cnt(v_4)$ accordingly. The cells involving a node computation are marked grey. Totally 2 iterations are needed. The four nodes $v_3$, $v_4$, $v_5$, and $v_6$ with $\status$ being $\statusy$ have their core numbers updated. Compared to \refex{semiadd}, we decrease the number of node computations from $12$ to $5$. 
\end{example}

\vspace*{-0.2cm}
\section{Performance Studies}
\label{sec:experiment}
In this section, we experimentally evaluate the performance of our proposed algorithms for both core decomposition and core maintenance. \refsubsec{performance} compares our solutions with state-of-the-art algorithms; \refsubsec{dynamic} shows the efficiency of our maintenance algorithm; and we reports the algorithm scalability in \refsubsec{scal}. 

All algorithms are implemented in C++, using gcc complier at -O3 optimization level. All the experiments are performed under a Linux operating system running on a machine with an Intel Xeon 3.4GHz CPU, 16GB RAM and 7200 RPM SATA Hard Drives (2TB). The time cost of algorithms are measured as the amount of wall-clock time elapsed during the program's execution. We adhere to standard external memory model for I/O statistics \cite{Aggarwal1988}. 

\stitle{Datasets.} We use two groups of datasets to demonstrate the efficiency of our semi-external algorithm. Group one consists of six graphs with relatively smaller size: \dblp, \ytb, \wiki, \cpt, \lj and \orkut. Group two consists of six big graphs: \webbase, \itc, \twitter, \sk, \uk and \clueweb. The detailed information for the $12$ datasets is displayed in \reftab{datasets}.

\begin{table}[t!]
\begin{center}
{\footnotesize
\begin{tabular}{l|l|l|l|l}
\hline
{\bf Datasets} & $|V|$ & $|E|$ & density & $\kmax$ \\\hline\hline
\dblp & 317,080 & 1,049,866 & 3.31 & 113 \\
\ytb & 1,134,890 & 2,987,624 & 2.63 & 51 \\
\wiki & 2,394,385 & 5,021,410 & 2.10 & 131 \\
\cpt & 3,774,768 & 16,518,948 & 4.38 & 64 \\
\lj & 3,997,962 & 34,681,189 & 8.67 & 360 \\
\orkut & 3,072,441 & 117,185,083 & 38.14 & 253 \\\hline
\hline
\webbase & 118,142,155 & 1,019,903,190 & 8.63 & 1506 \\
\itc & 41,291,594 & 1,150,725,436 & 27.86 & 3224 \\
\twitter & 41,652,230 & 1,468,365,182 & 35.25 & 2488 \\
\sk & 50,636,154 & 1,949,412,601 & 38.49 & 4510 \\
\uk & 105,896,555 & 3,738,733,648 & 35.30 & 5704 \\
\clueweb & 978,408,098 & 42,574,107,469 & 43.51 & 4244 \\\hline
\end{tabular}
}
\end{center}
\vspace*{-0.3cm}
\topcaption{Datasets} \label{tab:datasets}
\end{table}

In group one (small graphs), \dblp is a co-authorship network of the computer science bibliography DBLP. \ytb is a social network based on the user friendship in Youtube. \wiki is a network containing all the users and discussion from the inception of Wikipedia till January 2008. \cpt is citation graph includes all citations made by patents granted between 1975 and 1999. \lj (LiveJournal) is a free online blogging community where users declare friendships of each other. \orkut is a free online social network.

In group two (big graphs), \webbase is a graph obtained from the 2001 crawl performed by the WebBase crawler. \itc is a fairly large crawl of the \url{.it} domain. \twitter is a social network collected from Twitter where nodes are users and edges follow tweet transmission. \sk is a graph obtained from a 2005 crawl of the \url{.sk} domain. \uk is a graph gathering a snapshot of about $100$ million pages for the DELIS project in May 2007. Finally, \clueweb is a web graph underlying the ClueWeb12 dataset. All datasets can be downloaded from SNAP\footnote{\small \url{http://snap.stanford.edu/index.html}} and LAW\footnote{\small \url{http://law.di.unimi.it/index.php}}.

\vspace*{-0.2cm}
\subsection{Core Decomposition}
\label{subsec:performance}

\stitle{Small Graphs.} To explicitly reveal the performance of our core decomposition algorithms, we select the external-memory core decomposition algorithm $\emcore$ \cite{Cheng2011} and the classical in-memory algorithm \cite{Batagelj2003}, denoted by $\imcore$ for comparison. 

\begin{figure}[t]
\begin{center}
\begin{tabular}[t]{c}
\hspace{-1em}
\subfigure[Time Cost (Small Graphs)]{
	\includegraphics[width=0.5\columnwidth]{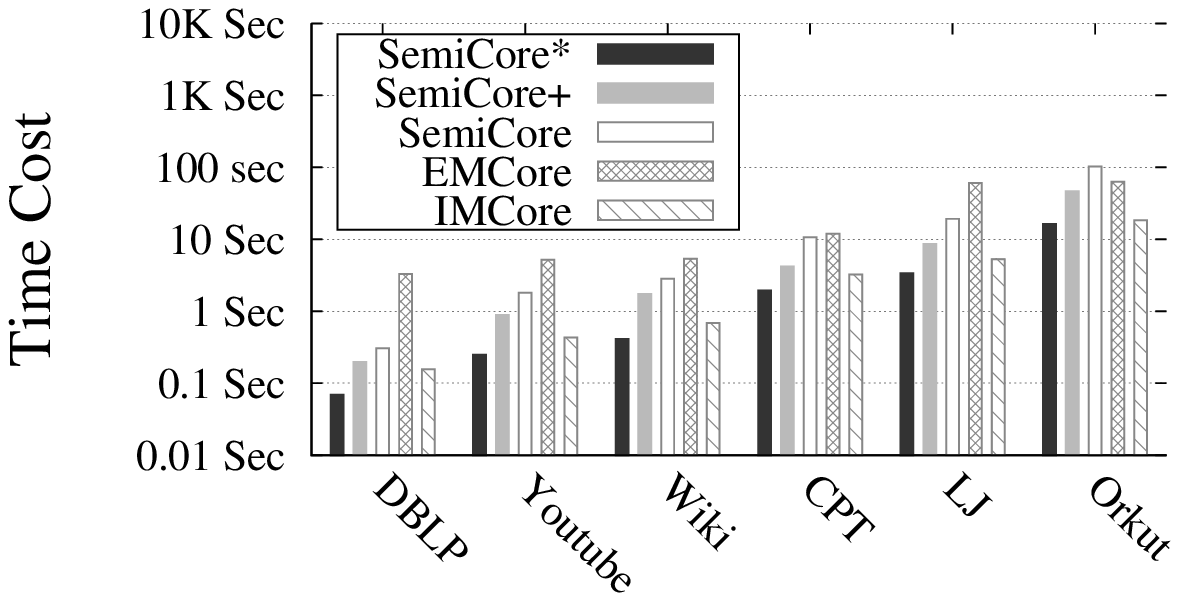}
}
\hspace{-1em}
\subfigure[Time Cost (Big Graphs)]{
	\includegraphics[width=0.5\columnwidth]{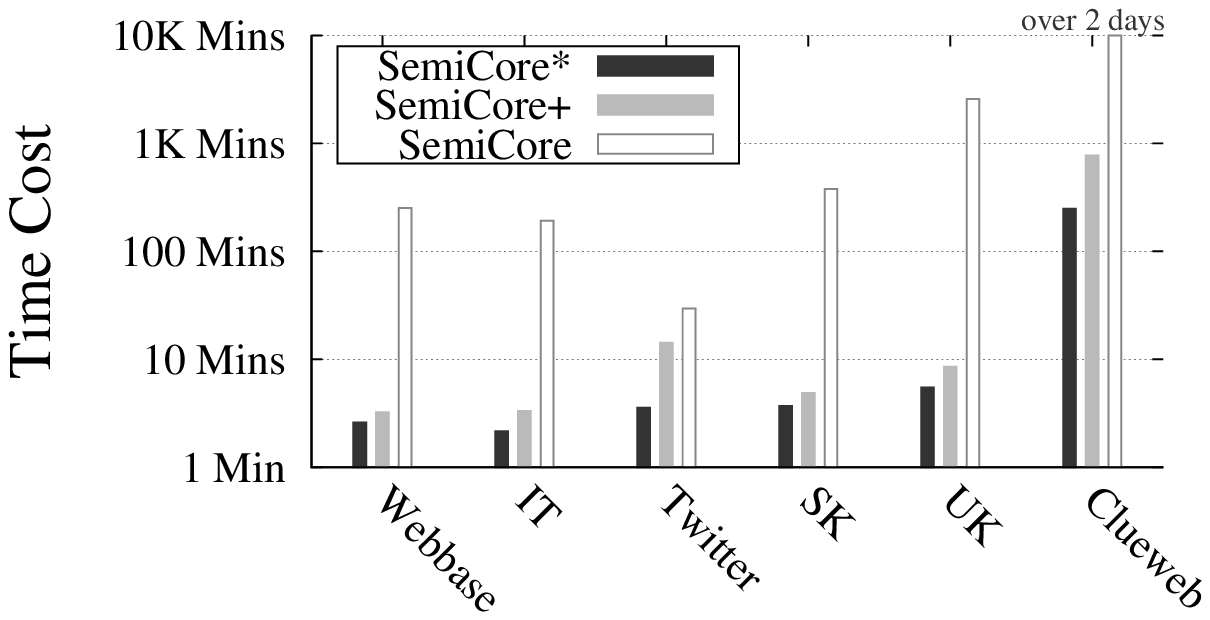}
}
\vspace*{-0.3cm}\\
\hspace{-1em}
\subfigure[Memory Usage (Small Graphs)]{
	\includegraphics[width=0.5\columnwidth]{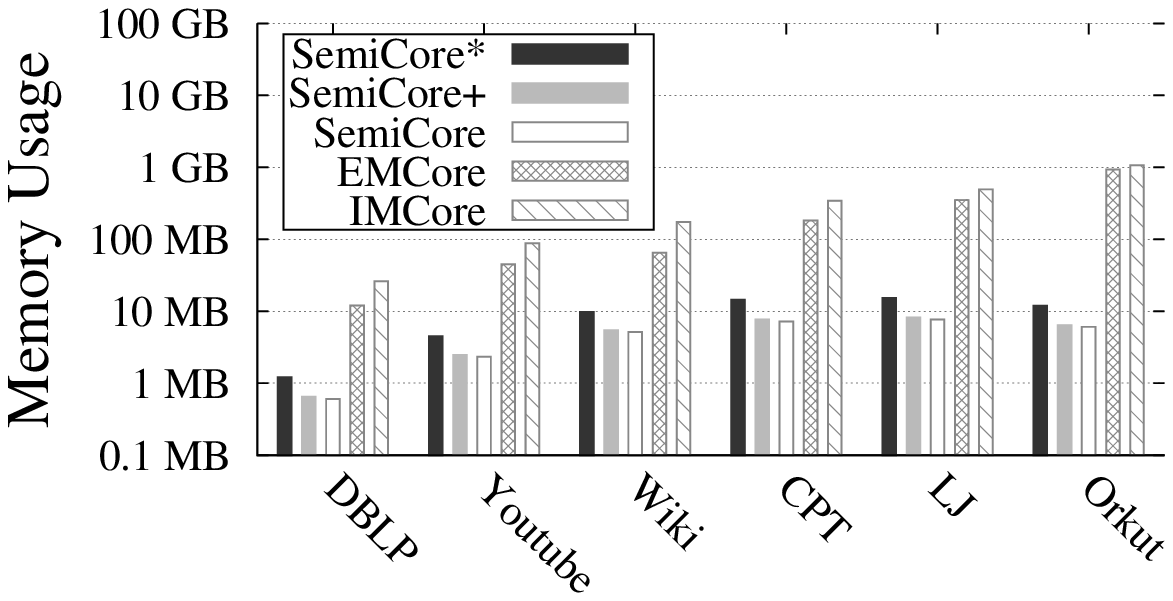}
}
\hspace{-1em}
\subfigure[Memory Usage (Big Graphs)]{
	\includegraphics[width=0.5\columnwidth]{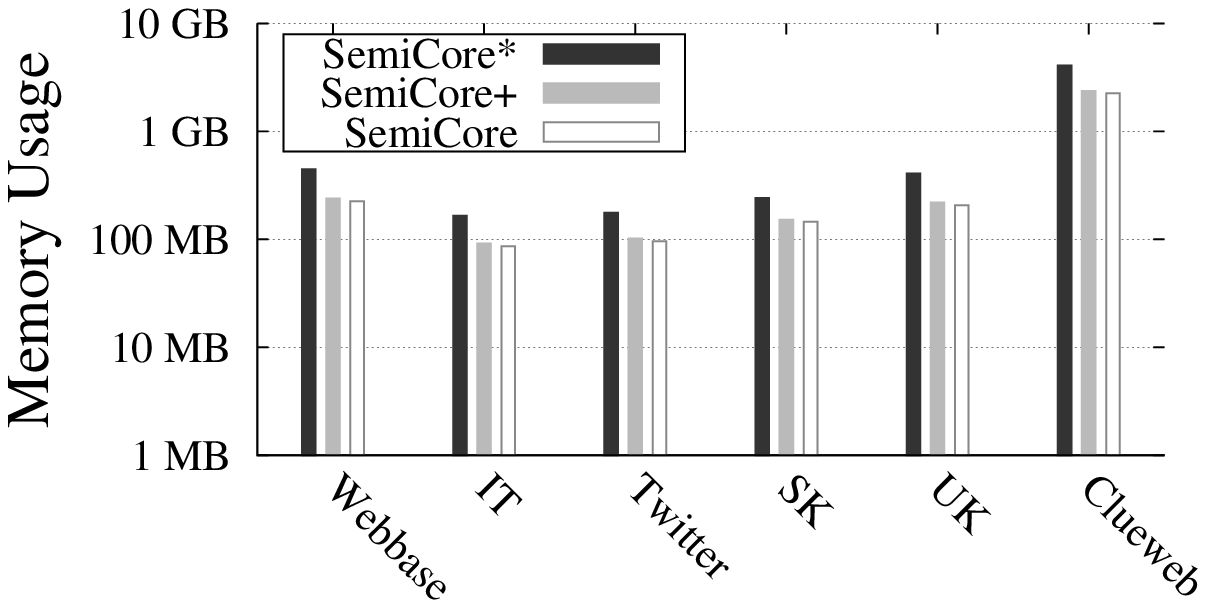}
}
\vspace*{-0.3cm}\\
\hspace{-1em}
\subfigure[I/Os (Small Graphs)]{
	\includegraphics[width=0.5\columnwidth]{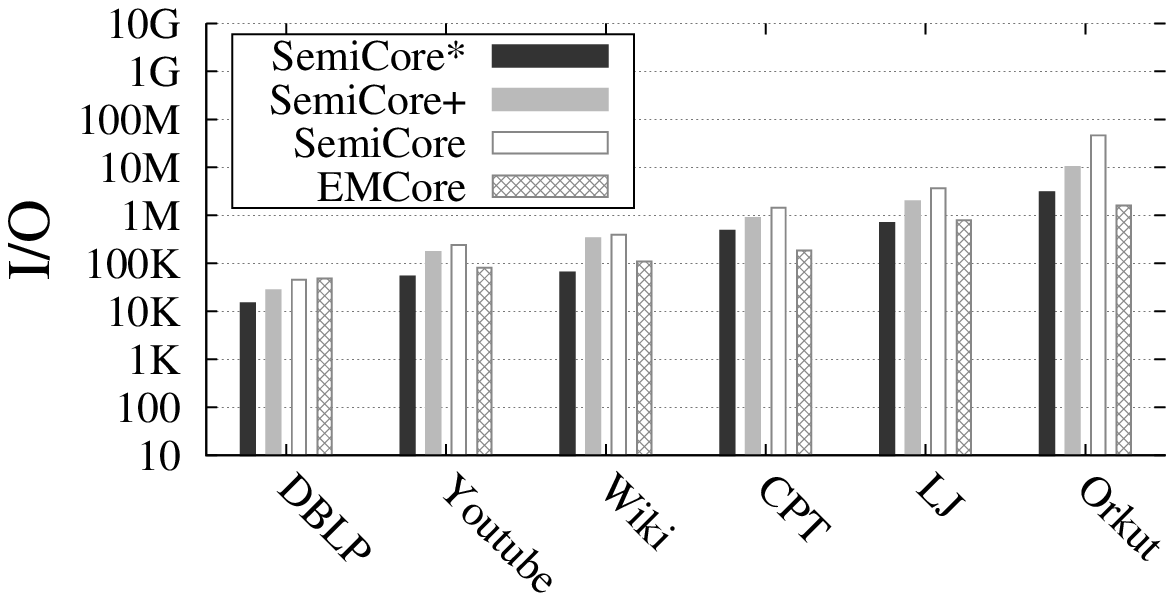}
}
\hspace{-1em}
\subfigure[I/Os (Big Graphs)]{
	\includegraphics[width=0.5\columnwidth]{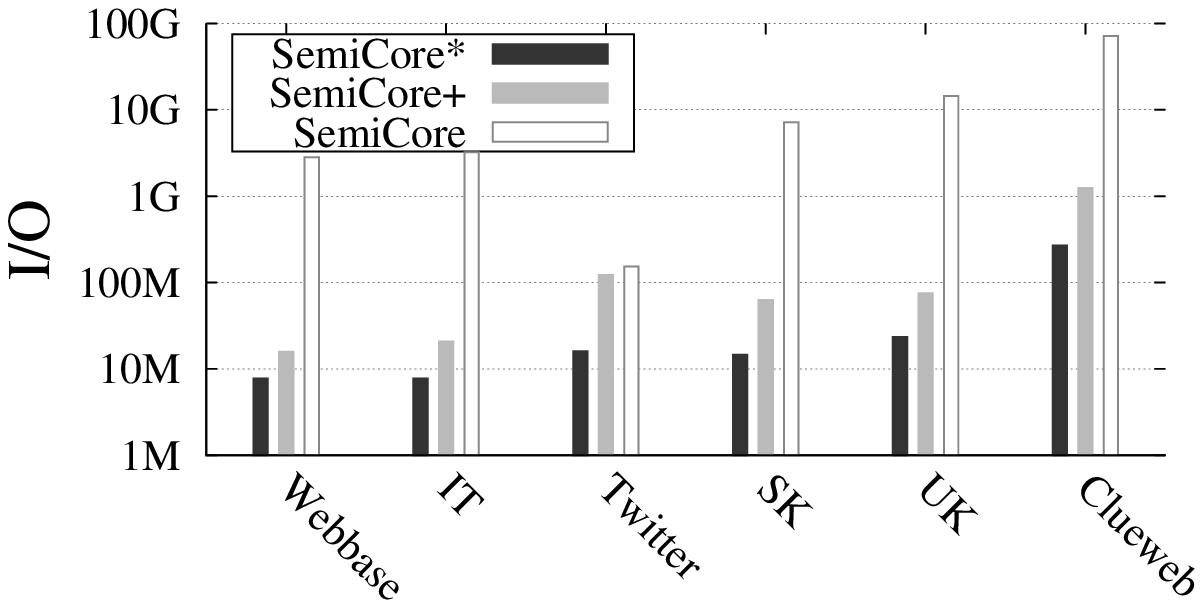}
}
\end{tabular}
\end{center}
\vspace*{-0.4cm}
\topcaption{Core Decomposition on Different Datasets}
\vspace*{-0.4cm}
\label{fig:exp:performance}
\end{figure}

As shown in \reffig{exp:performance} (a), the total running time of $\semicores$ is $10$ times faster than that of the $\emcore$ on average. It is remarkable that $\semicores$ can be even faster than the in-memory algorithm $\imcore$.  \reffig{exp:performance} (c) shows that algorithm $\semicores$ requires less memory than $\emcore$ and $\imcore$. Among all algorithms, $\semicore$ uses least memory since it does not rely on the \cnt numbers for all nodes comparing to $\semicores$. By contrast, $\emcore$ consumes a large amount of memory. Especially in  \orkut and \cpt, $\emcore$ consumes almost the same memory size as $\imcore$.  \reffig{exp:performance} (e) shows the I/O consumption of all algorithms except $\imcore$. $\semicores$ and $\emcore$ usually consume the least I/Os. However, due to the simple read-only data access of $\semicores$, $\semicores$ is much more efficient than $\emcore$ (refer to \reffig{exp:performance} (a)). 


\begin{figure}[t]
\begin{center}
\begin{tabular}[t]{c}
\hspace{-1em}
\subfigure[Average Time (Small Graphs)]{
	\includegraphics[width=0.5\columnwidth]{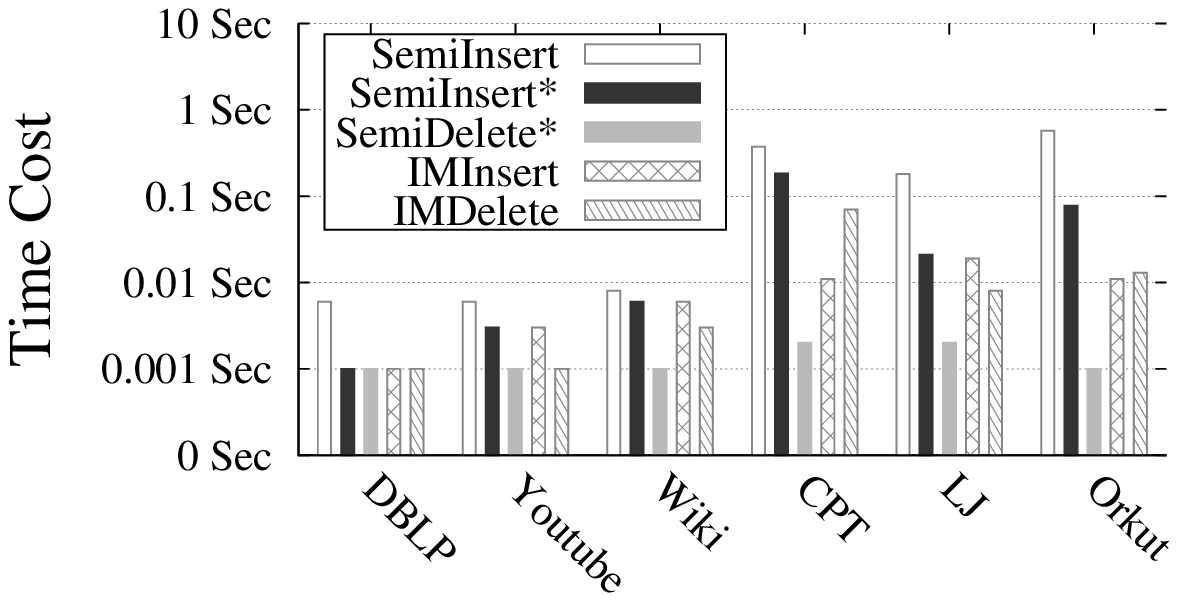}
}
\hspace{-1em}
\subfigure[Average Time (Big Graphs)]{
	\includegraphics[width=0.5\columnwidth]{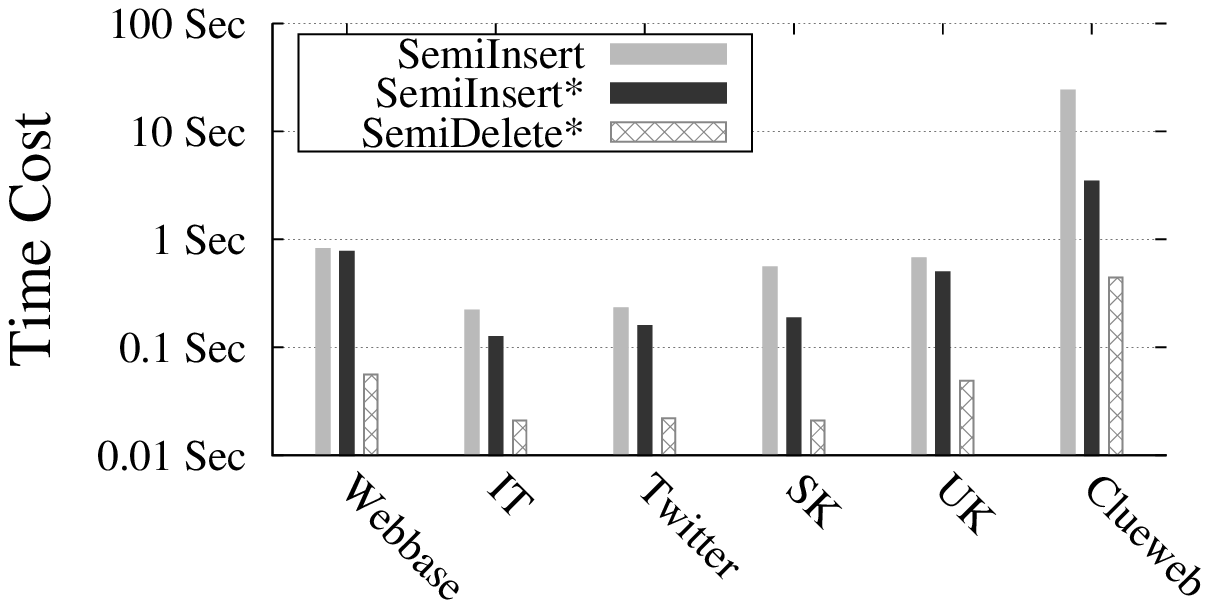}
}
\vspace*{-0.3cm}\\
\hspace{-1em}
\subfigure[Average I/Os (Small Graphs)]{
	\includegraphics[width=0.5\columnwidth]{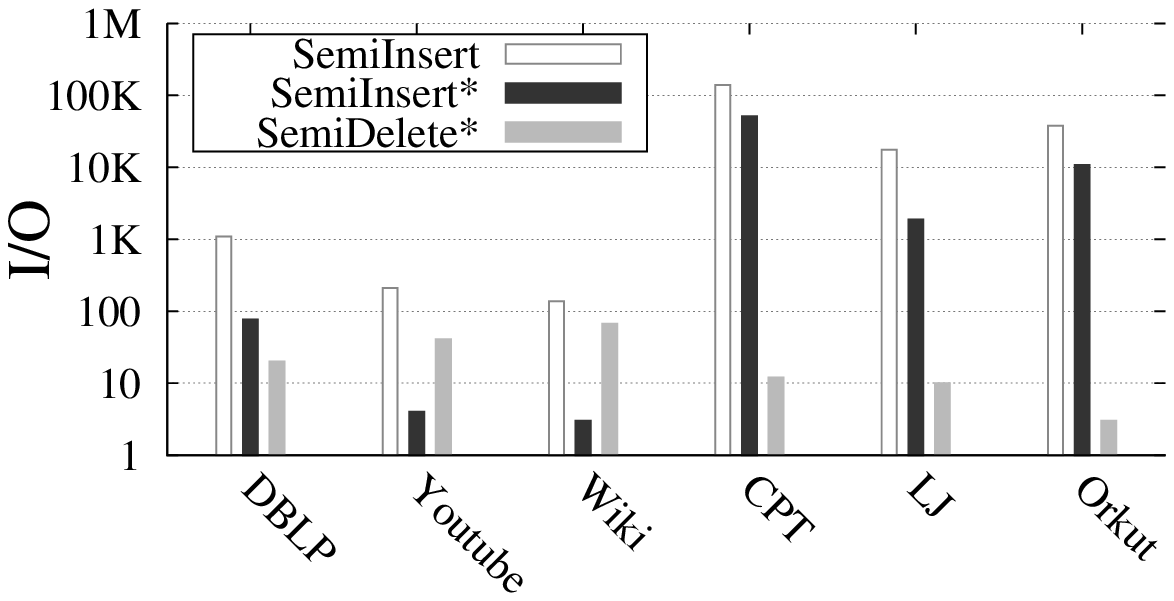}
}
\hspace{-1em}
\subfigure[Average I/Os (Big Graphs)]{
	\includegraphics[width=0.5\columnwidth]{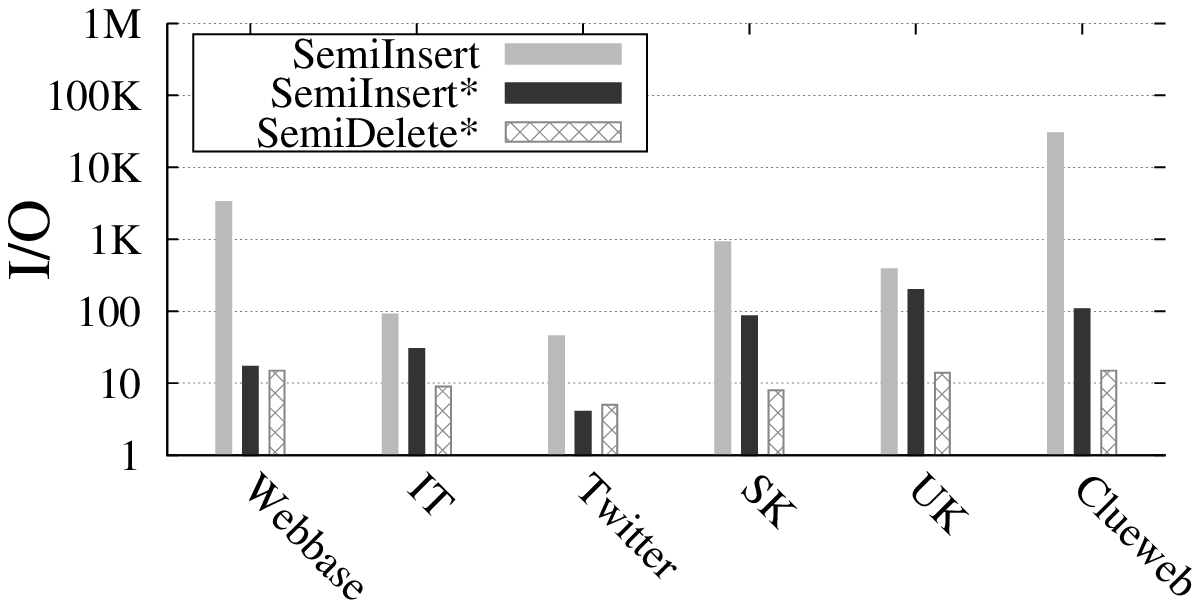}
}
\end{tabular}
\end{center}
\vspace*{-0.4cm}
\topcaption{Core Maintenance on Different Datasets}
\label{fig:exp:update}
\end{figure}

\stitle{Big Graphs.} We report the performance of our algorithms on big graphs in \reffig{exp:performance} (b), (d), and (f). The largest dataset \clueweb contains nearly $1$ billion nodes and $42.6$ billion edges. We can see from \reffig{exp:performance} (a) that $\semicores$ can process all datasets within $10$ minutes except \clueweb. In \reffig{exp:performance}, we can see that $\semicores$ totally costs less than $4.2$ GB memory  to process the largest dataset \clueweb. This result demonstrates that our algorithm can be deploy in any commercial machine to process big graph data. \reffig{exp:performance} (f) further reveals the advance of optimization in terms of I/O cost, since $\semicores$ spends much less I/Os than $\semicore$ and $\semicorep$ in all datasets.

\vspace*{-0.2cm}
\subsection{Core Maintenance}
\label{subsec:dynamic}

We test the performance of our maintenance algorithms ($\semiadd$, $\semiadds$, and $\semidels$). The state-of-the-art streaming in-memory algorithms in \cite{Sariyuce2013}, denoted by $\imadd$ and $\imdel$ are also compared in small graphs. We randomly select $100$ distinct existing edges in the graph for each test. To test the performance of edge deletion, we remove the $100$ edges from the graph one by one and take the average processing time and I/Os. To test the performance of edge insertion, after the $100$ edges are removed, we insert them into the graph one by one and take the average processing time and I/Os. The experimental results are reported in \reffig{exp:update}.

From \reffig{exp:update}, we can see that  $\semidels$ is more efficient than $\semiadds$ in both processing time and I/Os for all datasets. This is because  $\semidels$ simply follows $\semicores$ and does not rely on the calculation of other new graph properties. From \reffig{exp:update} (a), we can find that our core maintenance algorithm $\semiadds$ is comparable to the state-of-the-art in-memory algorithm $\imadd$ for edge insertion. $\semidels$ is even faster than $\imdel$ for edge deletion. This is due to the simple structures and data access model used in $\semidels$. $\semiadds$ outperforms $\semiadd$ in both processing time and I/Os for all datasets.

\begin{figure}[t]
\begin{center}
\begin{tabular}[t!]{c}
\hspace{-1em}
\subfigure[Vary $|V|$ (\twitter)]{
	\includegraphics[width=0.5\columnwidth]{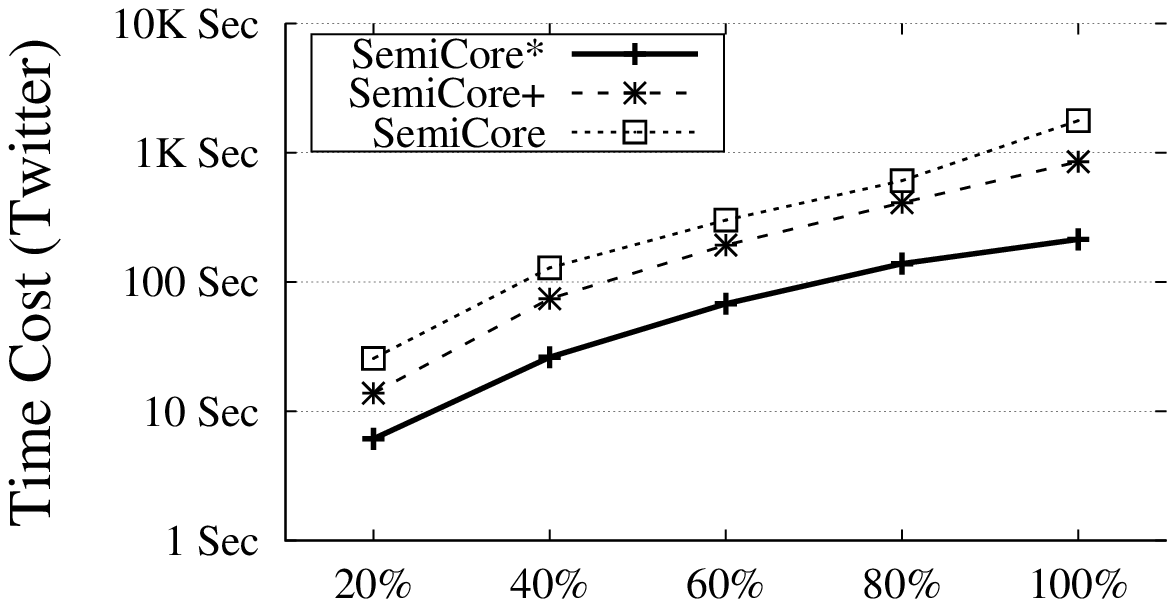}
}
\hspace{-1em}
\subfigure[Vary $|V|$ (\uk)]{
	\includegraphics[width=0.5\columnwidth]{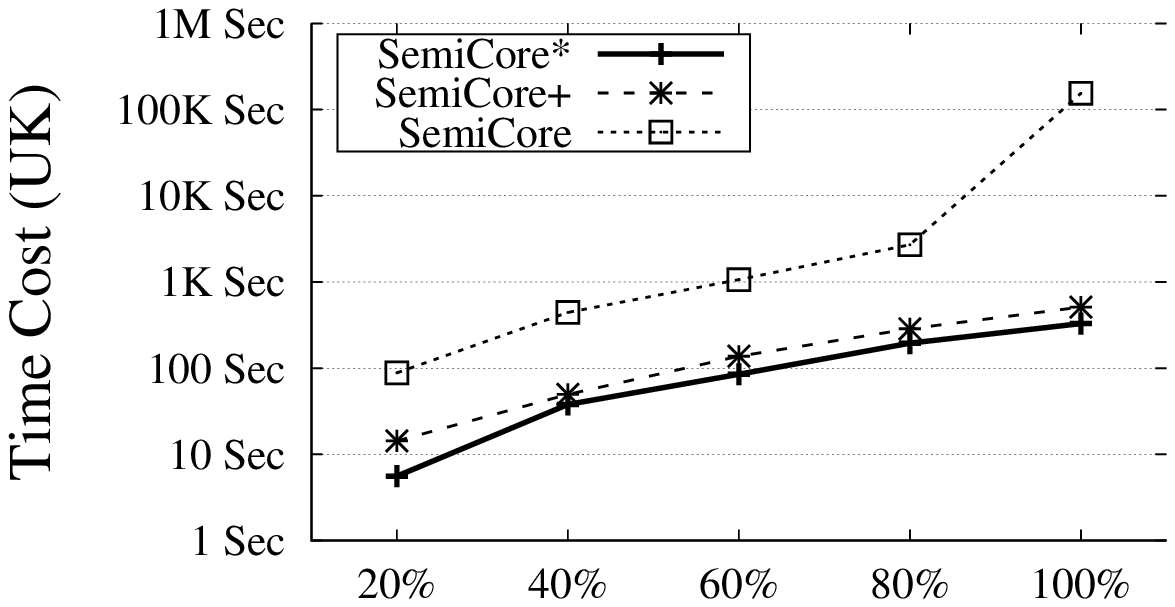}
}
\vspace*{-0.3cm}\\
\hspace{-1em}
\subfigure[Vary $|E|$ (\twitter)]{
	\includegraphics[width=0.5\columnwidth]{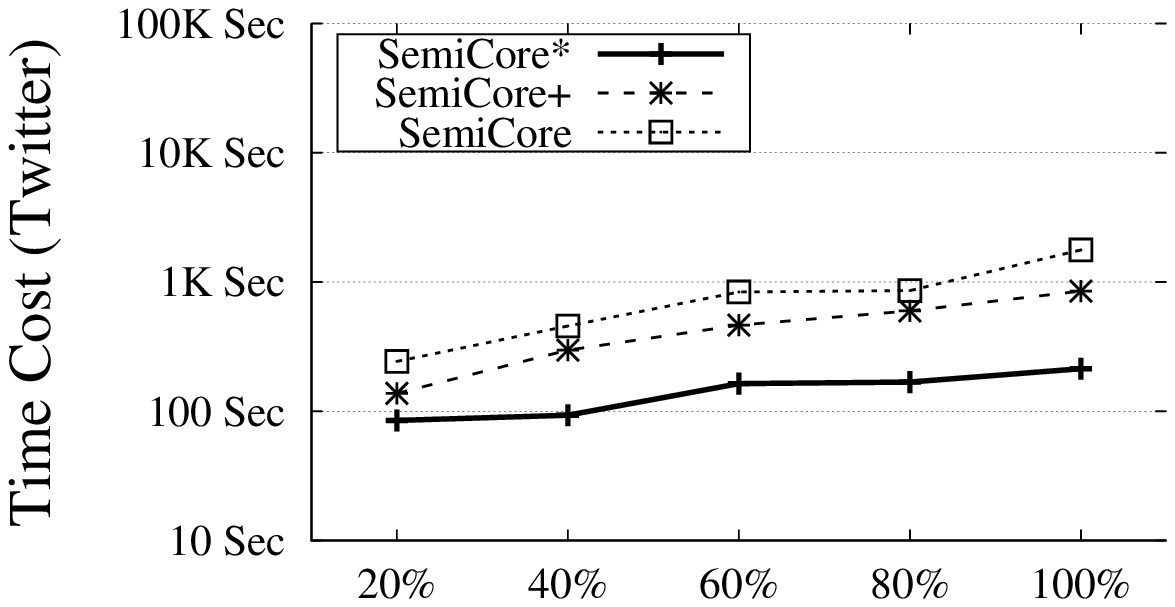}
}
\hspace{-1em}
\subfigure[Vary $|E|$ (\uk)]{
	\includegraphics[width=0.5\columnwidth]{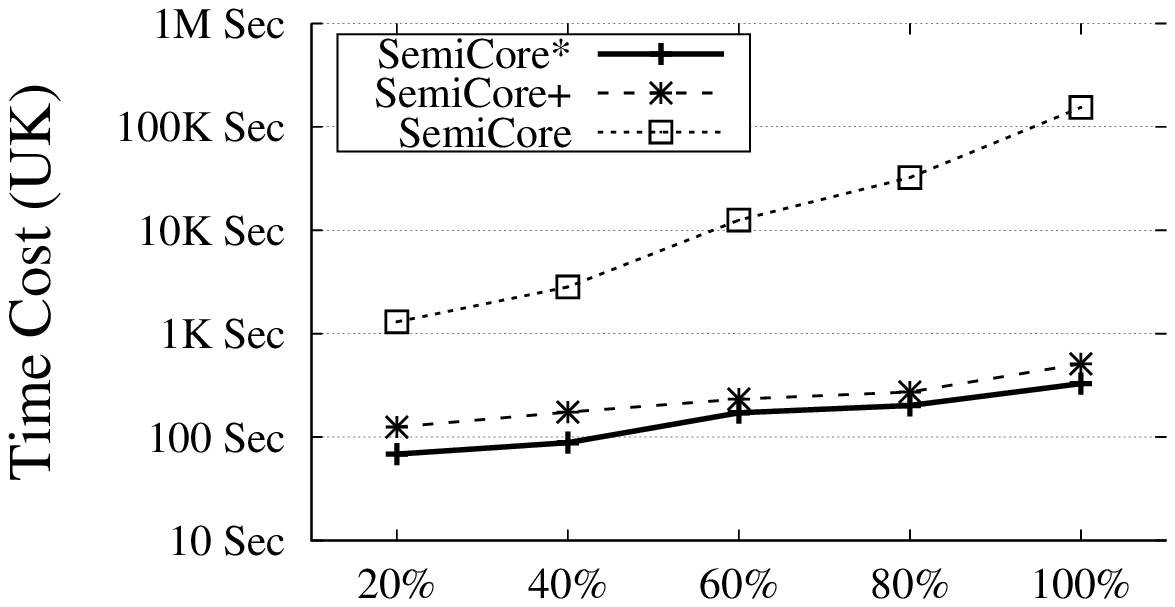}
}

\end{tabular}
\end{center}
\vspace*{-0.4cm}
\topcaption{Scalability of Core Decomposition}
\vspace*{-0.4cm}
\label{fig:exp:scal-static}
\end{figure}

\begin{figure}[t]
\begin{center}
\begin{tabular}[t!]{c}
\hspace{-1em}
\subfigure[Vary $|V|$ (\twitter)]{
	\includegraphics[width=0.5\columnwidth]{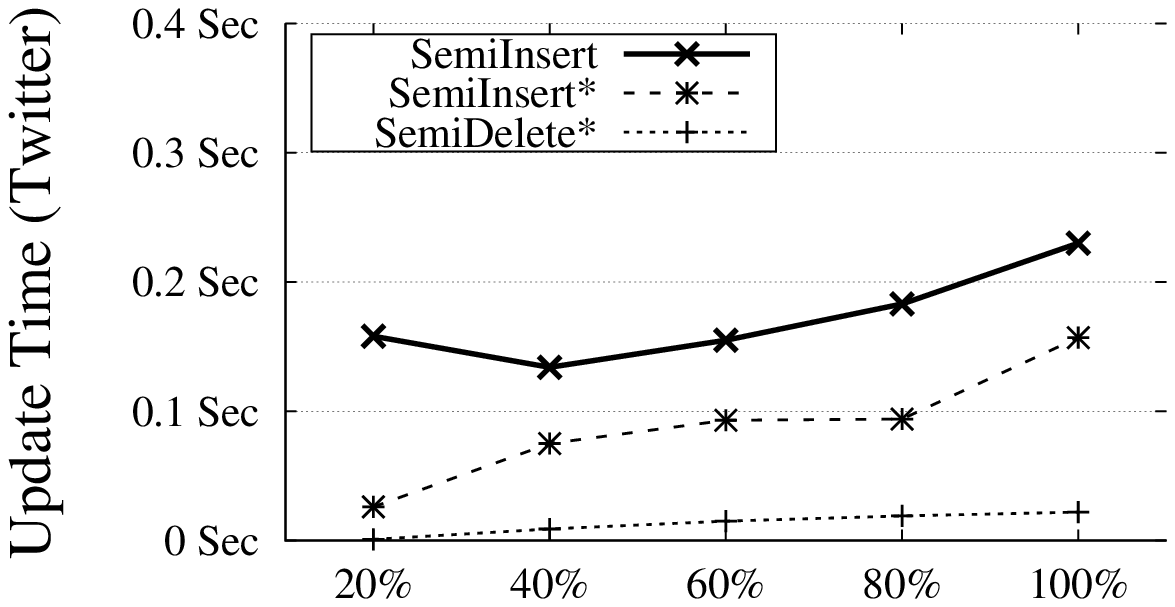}
}
\hspace{-1em}
\subfigure[Vary $|V|$ (\uk)]{
	\includegraphics[width=0.5\columnwidth]{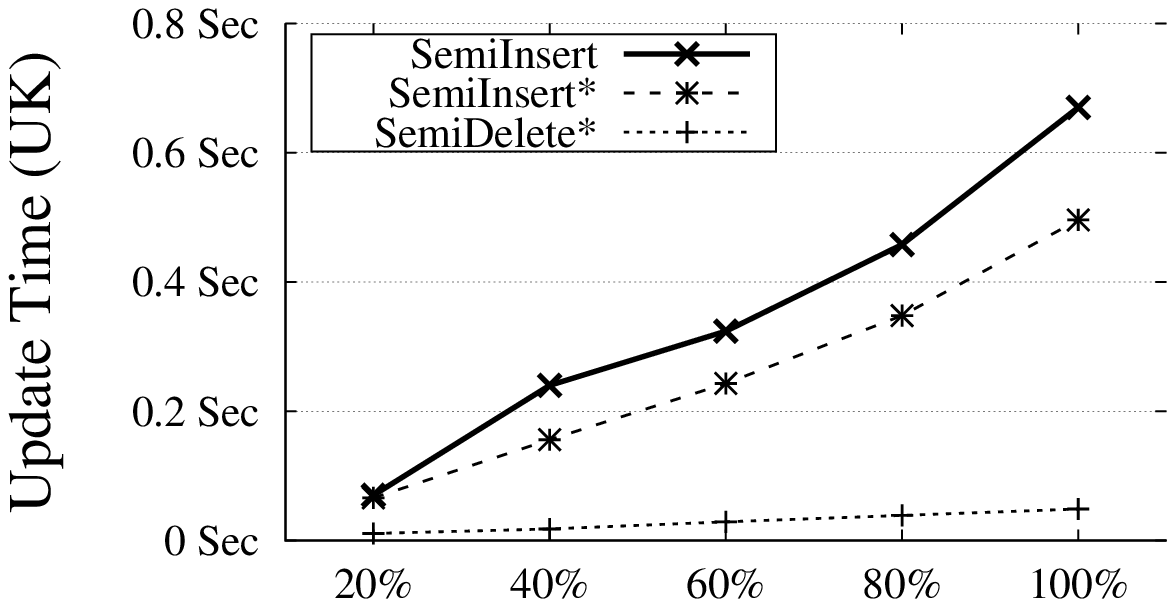}
}
\vspace*{-0.3cm}\\
\hspace{-1em}
\subfigure[Vary $|E|$ (\twitter)]{
	\includegraphics[width=0.5\columnwidth]{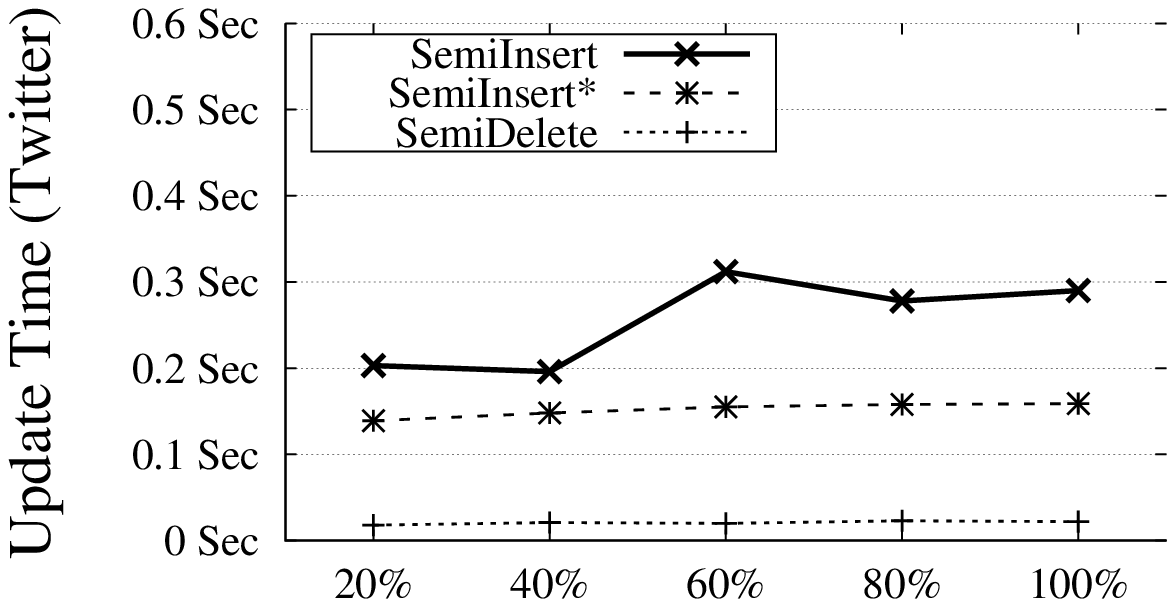}
}
\hspace{-1em}
\subfigure[Vary $|E|$ (\uk)]{
	\includegraphics[width=0.5\columnwidth]{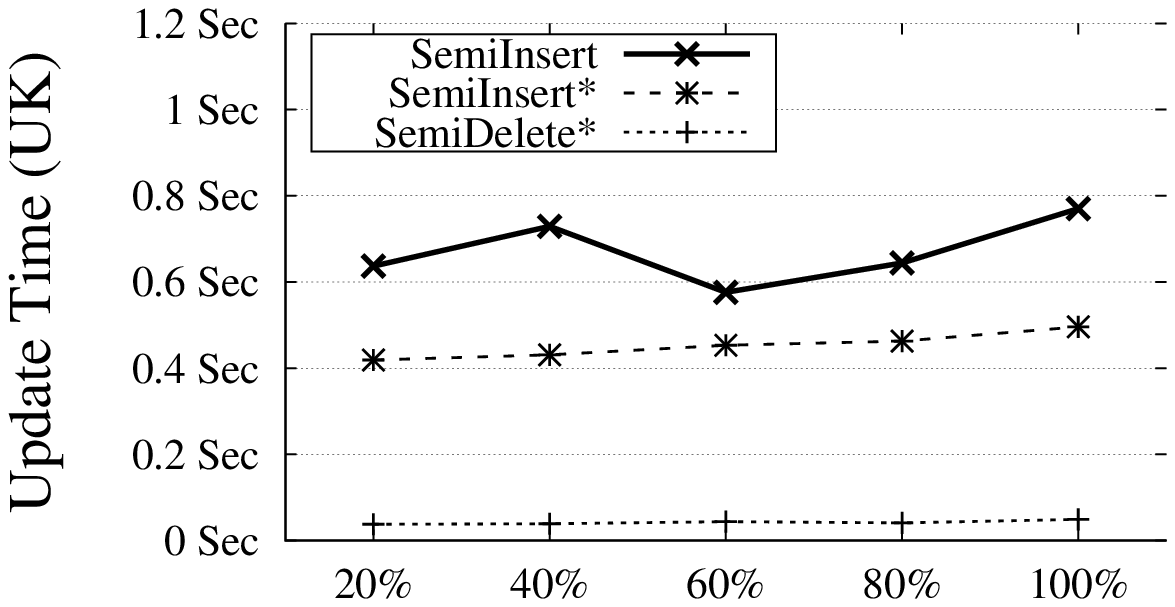}
}

\end{tabular}
\end{center}
\vspace*{-0.4cm}
\topcaption{Scalability of Core Maintenance }
\label{fig:exp:scal-dynamic}
\end{figure}

\vspace*{-0.2cm}
\subsection{Scalability Testing}
\label{subsec:scal}

In this experiment, we test the scalability of our core decomposition and core maintenance algorithms.  We choose two big graphs \twitter and \uk for testing. We vary number of nodes $|V|$ and number of edges $|E|$ of \twitter and \uk by randomly sampling nodes and edges respectively from 20\% to 100\%. When sampling nodes, we keep the induced subgraph of the nodes, and when sampling edges, we keep the incident nodes of the edges. Here, we only report the processing time. The memory usage is linear to the number of nodes, and the curves for I/O cost are similar to that of processing time. 

\stitle{Core Decomposition}. \reffig{exp:scal-static} (a) and (b) report the processing time of our proposed algorithms for core decomposition when varying $|V|$ in \twitter and \uk respectively. When $|V|$ increases, the processing time for all algorithms increases. $\semicores$ performs best in all cases and is over an order of magnitude faster than $\semicore$ in both \twitter and \uk. \reffig{exp:scal-static} (a) and (b) show the processing time of our core decomposition algorithms when varying $|E|$ in \twitter and \uk respectively. When $|E|$ increases, the processing time for all algorithms increases, and $\semicores$ performs best among all three algorithms. When $|E|$ increases, the gap between $\semicores$ and $\semicore$ also increases. For example, in \uk, when $|E|$ reaches $100\%$, $\semicores$ is more than two orders of magnitude faster than $\semicore$.


\stitle{Core Maintenance}. The scalability testing results for core maintenance are shown in \reffig{exp:scal-dynamic}. As shown in \reffig{exp:scal-dynamic} (a) and \reffig{exp:scal-dynamic} (b),  when increasing $|V|$ from $20\%$ to $100\%$, the processing time for all algorithms increases. $\semidels$ performs best, and $\semiadds$ is faster than $\semiadd$ for all testing cases. The curves of our core maintenance algorithms when varying $|E|$ are shown in \reffig{exp:scal-dynamic} (c) and \reffig{exp:scal-dynamic} (d) for \twitter and \uk respectively. $\semidels$ and $\semiadds$ are very stable when increasing $|E|$ in both \twitter and \uk, which shows the high scalability of our core maintenance algorithms. $\semiadd$ performs worst among all three algorithms. When $|E|$ increases, the performance of $\semiadd$ is unstable because $\semiadd$ needs to locate a connected component whose size can be very large in some cases.  


\section{Related Work}
\label{sec:relatedwork}
\stitle{Core Decomposition}. $k$-core is first introduced in \cite{seidman1983}. Batagelj and Zaversnik \cite{Batagelj2003} give an linear in-memory algorithm for core decomposition, which is presented detailed in \refsec{existing}. This problem is also studied for the weighted graphs \cite{Giatsidis2011} and directed graphs \cite{giatsidis2011d}.   Cheng et al. \cite{Cheng2011} propose an I/O efficient algorithm for core decomposition.  \cite{montresor2013} gives a distributed algorithm for core decomposition. Core decomposition in random graphs is studied in \cite{Luczak1991,Pittel1996,Molloy2005,Janson2007}. Core decomposition in an uncertain graph is  studied in \cite{BonchiGKV14}. Locally computing and estimating core numbers are studied in \cite{Cui2014} and \cite{OBrienS14} respectively.   \cite{Sariyuce2013} and  \cite{LiYM14} propose  in-memory algorithms to maintain the core numbers of nodes in dynamic graphs. 




\stitle{Semi-external Algorithms.} Semi-external model, which strictly bounds the memory size, becomes very popular in processing big graphs recently. For example, \cite{zhiwei2013} proposes a semi-external algorithm to find all strong connected components for a massive directed graph. \cite{zhiwei2015} gives semi-external algorithms to compute a DFS tree for a graph in the disk using divide \& conquer strategy. \cite{liu2015} studies maximum independent set under the semi-external model.


\section{Conclusions}
\label{sec:conclusion}
In this paper, considering that many real-world graphs are big and cannot reside in the main memory of a machine,  we study I/O efficient core decomposition on web-scale graphs, which has a large number of applications. The existing solution is not scalable to handle big graphs because it cannot bound the memory size and may load most part of the graph in memory. Therefore, we follow a semi-external model, which can well bound the memory size. We propose an I/O efficient semi-external algorithm for core decomposition, and explore two optimization strategies to further reduce the I/O and CPU cost. We further propose semi-external algorithms and optimization techniques to handle graph updates. We conduct extensive experiments on $12$ real graphs, one of which contains $978.5$ million nodes and $42.6$ billion edges, to demonstrate the efficiency of our proposed algorithm.


{\footnotesize
\bibliographystyle{abbrv}
\bibliography{scd}

\begin{thebibliography}{10}

\bibitem{Aggarwal1988}
A.~Aggarwal and S.~Vitter, Jeffrey.
\newblock The input/output complexity of sorting and related problems.
\newblock {\em Commun. ACM}, 31(9), 1988.

\bibitem{Altaf2003}
M.~Altaf-Ul-Amine, K.~Nishikata, T.~Korna, T.~Miyasato, Y.~Shinbo,
  M.~Arifuzzaman, C.~Wada, M.~Maeda, T.~Oshima, H.~Mori, and S.~Kanaya.
\newblock Prediction of protein functions based on k-cores of protein-protein
  interaction networks and amino acid sequences.
\newblock {\em Genome Informatics}, 14, 2003.

\bibitem{kalvarez2005}
J.~I. Alvarez{-}Hamelin, L.~Dall'Asta, A.~Barrat, and A.~Vespignani.
\newblock k-core decomposition: a tool for the visualization of large scale
  networks.
\newblock {\em CoRR}, abs/cs/0504107, 2005.

\bibitem{alvarez2005}
J.~I. Alvarez-Hamelin, L.~Dall'Asta, A.~Barrat, and A.~Vespignani.
\newblock Large scale networks fingerprinting and visualization using the
  k-core decomposition.
\newblock In {\em Proc. of NIPS'05}, 2005.

\bibitem{alvarez2006}
J.~I. Alvarez-Hamelin, L.~Dall’Asta, A.~Barrat, and A.~Vespignani.
\newblock How the k-core decomposition helps in understanding the internet
  topology.
\newblock In {\em ISMA Workshop on the Internet Topology}, volume~1, 2006.

\bibitem{andersen2009}
R.~Andersen and K.~Chellapilla.
\newblock Finding dense subgraphs with size bounds.
\newblock In {\em Algorithms and Models for the Web-Graph}. 2009.

\bibitem{bader2003}
G.~Bader and C.~Hogue.
\newblock An automated method for finding molecular complexes in large protein
  interaction networks.
\newblock {\em BMC Bioinformatics}, 4(1), 2003.

\bibitem{balasundaram2011}
B.~Balasundaram, S.~Butenko, and I.~V. Hicks.
\newblock Clique relaxations in social network analysis: The maximum k-plex
  problem.
\newblock {\em Operations Research}, 59(1), 2011.

\bibitem{Batagelj2003}
V.~Batagelj and M.~Zaversnik.
\newblock An o(m) algorithm for cores decomposition of networks.
\newblock {\em CoRR}, cs.DS/0310049, 2003.

\bibitem{BonchiGKV14}
F.~Bonchi, F.~Gullo, A.~Kaltenbrunner, and Y.~Volkovich.
\newblock Core decomposition of uncertain graphs.
\newblock In {\em Proc. of KDD'14}, 2014.

\bibitem{Cheng2011}
J.~Cheng, Y.~Ke, S.~Chu, and M.~T. {\"O}zsu.
\newblock Efficient core decomposition in massive networks.
\newblock In {\em Proc. of ICDE'11}, 2011.

\bibitem{Cui2014}
W.~Cui, Y.~Xiao, H.~Wang, and W.~Wang.
\newblock Local search of communities in large graphs.
\newblock In {\em Proc. of SIGMOD'14}, 2014.

\bibitem{dorogovtsev2006}
S.~N. Dorogovtsev, A.~V. Goltsev, and J.~F.~F. Mendes.
\newblock K-core organization of complex networks.
\newblock {\em Physical review letters}, 96(4), 2006.

\bibitem{giatsidis2011d}
C.~Giatsidis, D.~M. Thilikos, and M.~Vazirgiannis.
\newblock D-cores: Measuring collaboration of directed graphs based on
  degeneracy.
\newblock In {\em Proc. of ICDM'11}, 2011.

\bibitem{Giatsidis2011}
C.~Giatsidis, D.~M. Thilikos, and M.~Vazirgiannis.
\newblock Evaluating cooperation in communities with the k-core structure.
\newblock In {\em Proc. of ASONAM'11}, 2011.

\bibitem{healy2008}
J.~Healy, J.~Janssen, E.~Milios, and W.~Aiello.
\newblock Characterization of graphs using degree cores.
\newblock In {\em Algorithms and Models for the Web-Graph}. 2008.

\bibitem{Janson2007}
S.~Janson and M.~J. Luczak.
\newblock A simple solution to the k-core problem.
\newblock {\em Random Struct. Algorithms}, 30(1-2), 2007.

\bibitem{LiQYM15}
R.~Li, L.~Qin, J.~X. Yu, and R.~Mao.
\newblock Influential community search in large networks.
\newblock {\em {PVLDB}}, 8(5), 2015.

\bibitem{LiYM14}
R.~Li, J.~X. Yu, and R.~Mao.
\newblock Efficient core maintenance in large dynamic graphs.
\newblock {\em {IEEE} Trans. Knowl. Data Eng.}, 26(10), 2014.

\bibitem{liu2015}
Y.~Liu, J.~Lu, H.~Yang, X.~Xiao, and Z.~Wei.
\newblock Towards maximum independent sets on massive graphs.
\newblock {\em PVLDB}, 8(13), 2015.

\bibitem{Luczak1991}
T.~Luczak.
\newblock Size and connectivity of the k-core of a random graph.
\newblock {\em Discrete Math.}, 91(1), 1991.

\bibitem{Molloy2005}
M.~Molloy.
\newblock Cores in random hypergraphs and boolean formulas.
\newblock {\em Random Struct. Algorithms}, 27(1), 2005.

\bibitem{montresor2013}
A.~Montresor, F.~De~Pellegrini, and D.~Miorandi.
\newblock Distributed k-core decomposition.
\newblock {\em TPDS}, 24(2), 2013.

\bibitem{OBrienS14}
M.~P. O'Brien and B.~D. Sullivan.
\newblock Locally estimating core numbers.
\newblock In {\em Proc. of ICDM'14}, 2014.

\bibitem{Pittel1996}
B.~Pittel, J.~Spencer, and N.~Wormald.
\newblock Sudden emergence of a giant k-core in a random graph.
\newblock {\em J. Comb. Theory Ser. B}, 67(1), 1996.

\bibitem{QinLCZ15}
L.~Qin, R.~Li, L.~Chang, and C.~Zhang.
\newblock Locally densest subgraph discovery.
\newblock In {\em Proc. of KDD'15}, 2015.

\bibitem{Sariyuce2013}
A.~E. Sar\'{\i}y\"{u}ce, B.~Gedik, G.~Jacques-Silva, K.-L. Wu, and U.~V.
  \c{C}ataly\"{u}rek.
\newblock Streaming algorithms for k-core decomposition.
\newblock {\em PVLDB}, 6(6), 2013.

\bibitem{seidman1983}
S.~B. Seidman.
\newblock Network structure and minimum degree.
\newblock {\em Social networks}, 5(3), 1983.

\bibitem{SozioG10}
M.~Sozio and A.~Gionis.
\newblock The community-search problem and how to plan a successful cocktail
  party.
\newblock In {\em Proc. of KDD'10}, 2010.

\bibitem{verma2012}
A.~Verma and S.~Butenko.
\newblock Network clustering via clique relaxations: A community based
  approach.
\newblock {\em Graph Partitioning and Graph Clustering}, 588, 2012.

\bibitem{zhang2010}
H.~Zhang, H.~Zhao, W.~Cai, J.~Liu, and W.~Zhou.
\newblock Using the k-core decomposition to analyze the static structure of
  large-scale software systems.
\newblock {\em The Journal of Supercomputing}, 53(2), 2010.

\bibitem{zhiwei2013}
Z.~Zhang, J.~X. Yu, L.~Qin, L.~Chang, and X.~Lin.
\newblock {I/O} efficient: computing sccs in massive graphs.
\newblock In {\em Proc. of SIGMOD'13}, 2013.

\bibitem{zhiwei2015}
Z.~Zhang, J.~X. Yu, L.~Qin, and Z.~Shang.
\newblock Divide {\&} conquer: {I/O} efficient depth-first search.
\newblock In {\em Proc. of SIGMOD'15}, 2015.

\end{thebibliography}
}

\end{document}